\documentclass[conference]{IEEEtran}

\usepackage[english]{babel}
\usepackage{amsmath}
\usepackage{graphicx}
\usepackage{subfig}
\usepackage{tabularx}
\usepackage{tikz}
\usepackage{multirow}
\usepackage{xspace}
\usepackage{xurl}
\usepackage[most]{tcolorbox}
\usepackage{adjustbox}
\usepackage{longtable}
\usepackage[framemethod=TikZ]{mdframed}

\usepackage{shortcuts}

\newtcbtheorem{Summary}{\bfseries Takeaway}{enhanced,drop shadow={black!50!white},
  coltitle=black,
  top=0.07in,
      bottom=1pt,
  attach boxed title to top left=
    {xshift=1.5em,yshift=-\tcboxedtitleheight/2},
  boxed title style={size=small,colback=pink}
}{summary}

\newtcolorbox[auto counter]{summary}[1][]{title={\bfseries Summary~\thetcbcounter},enhanced,drop shadow={black!50!white},
  coltitle=black,
  top=0.1in,
  attach boxed title to top left=
  {xshift=1.5em,yshift=-\tcboxedtitleheight/2},
  boxed title style={size=small,colback=pink},#1}

\renewcommand{\paragraph}{\vspace{3pt}\noindent\textbf}
\usepackage{hyperref}  \usepackage{cleveref}

\usepackage[absolute,showboxes]{textpos}

\begin{document}

\date{}

\setlength{\TPHorizModule}{\paperwidth}
\setlength{\TPVertModule}{\paperheight}
\TPMargin{2pt}
\begin{textblock}{0.8}(0.1,0.02)
     \noindent
     \footnotesize
     \textcolor{blue!80!black}{If you cite this paper, please use the next reference:
     Gibran Gomez, Kevin Van Liebergen, Davide Sanvito, Giuseppe Siracusano, Roberto Gonzalez, and Juan Caballero.
     Clean Up the Mess: Addressing Data Pollution in Cryptocurrency Abuse Reporting Services.
     In \textit{Future Generation Computer Systems} (2025).
     DOI: \url{https://doi.org/10.1016/j.future.2025.108313}.}
\end{textblock}

\title{Clean Up the Mess: Addressing Data Pollution in\\ Cryptocurrency Abuse Reporting Services}

\author{
\IEEEauthorblockN{
Gibran Gomez\IEEEauthorrefmark{1}\IEEEauthorrefmark{2},
Kevin van Liebergen\IEEEauthorrefmark{1}\IEEEauthorrefmark{2},
Davide Sanvito\IEEEauthorrefmark{3},\\
Giuseppe Siracusano\IEEEauthorrefmark{3},
Roberto Gonzalez\IEEEauthorrefmark{3} and
Juan Caballero\IEEEauthorrefmark{1}
}
\\
\IEEEauthorblockA{\IEEEauthorrefmark{1}IMDEA Software Institute}
\IEEEauthorblockA{\IEEEauthorrefmark{2}Universidad Polit\'ecnica de Madrid}
\IEEEauthorblockA{\IEEEauthorrefmark{3}NEC Laboratories Europe}
}

\maketitle

\begin{abstract}

Cryptocurrency abuse reporting services are a valuable data source about 
abusive blockchain addresses, prevalent types of cryptocurrency abuse, and 
their financial impact on victims.
However, they may suffer data pollution due to their crowd-sourced nature.
\updated{This work analyzes the extent and impact
of data pollution in cryptocurrency abuse reporting services and 
proposes a novel LLM-based defense to address the pollution.} 
We collect 289K abuse reports submitted over 6 years to  
two popular services and 
use them to answer three research questions.
RQ1 analyzes the extent and impact of pollution. 
We show that spam reports will eventually flood unchecked abuse reporting services,
with \ba receiving 75\% of spam before stopping operations.
We build a public dataset of \numgtfinal abuse reports
labeled with 19 popular abuse types
and use it to reveal the inaccuracy of user-reported abuse types.
We identified 91 (0.1\%) benign addresses reported,
responsible for 60\% of all the received funds.
RQ2 examines whether we can automate identifying valid reports
and their classification into abuse types.
We propose an unsupervised LLM-based classifier that 
achieves an F1 score of 0.95 when classifying reports,
an F1 of 0.89 when classifying out-of-distribution data,
and an F1 of 0.99 when identifying spam reports.
\updated{Our unsupervised LLM-based classifier clearly outperforms two baselines:
a supervised classifier and a naive usage of the LLM.}
\updated{Finally, RQ3 demonstrates the usefulness of our LLM-based classifier 
for quantifying the financial impact of different cryptocurrency abuse types.}
We show that victim-reported losses
heavily underestimate cybercriminal revenue
by estimating a 29 times higher revenue
from deposit transactions.
We identified that investment scams have the highest financial impact 
and that extortions have lower conversion rates
but compensate for them with massive email campaigns.  

\end{abstract}

\section{Introduction}
\label{sec:intro}

Cryptocurrencies have become a popular payment and financial investment 
mechanism. 
Unfortunately, they are also frequently abused by cybercriminals,
for example, to receive payments from victims of 
criminal activities such as
ransomware~\cite{bitiodineSpagnoulo,behindLiao,economicConti,trackingHuang,ransomwareClouston},
sextortion~\cite{spamsPaquetClouston,oggier2020ego}, and
scams~\cite{ponziBartoletti,fistfulMeiklejohn,xia2020characterizing,bartoletti2021cryptocurrency,li2023giveaway}.
A valuable source of data on cryptocurrency abuse is reporting 
services that provide online forms for users to fill out 
with information about the abuse they have observed
(e.g., date, description, type, amount lost),
the abuser (e.g., email address, cryptocurrency address), and
the reporters themselves (e.g., country, age).
Some abuse reporting services specialize in cryptocurrency abuse
(e.g., BitcoinAbuse~\cite{bitcoinabuse}, ChainAbuse~\cite{chainabuse}),
while others collect all abuse reports,
regardless of whether they involve cryptocurrencies, such as 
the Better Business Bureau (BBB) Scam Tracker~\cite{scamtracker},
the Australian Competition and Consumer Commission (ACCC) 
ScamWatch~\cite{scamwatch}, and
the Federal Trade Commission (FTC) ReportFraud~\cite{reportfraud}.

Cryptocurrency abuse reports provide critical information 
on abusive cryptocurrency addresses, prevalent cryptocurrency abuse types, and 
financial losses suffered by victims.
Blockchain explorers~\cite{glasschain,crystalblockchain}
link cryptocurrency addresses to their abuse reports 
so that users can avoid known malicious addresses. 
The reported malicious cryptocurrency addresses are also invaluable 
for monitoring cryptocurrency transactions and reporting suspicious ones, 
as mandated by anti-money laundering (AML) requirements on 
cryptocurrency exchanges introduced by  
the US Bank Secrecy Act~\cite{bsa} and 
the EU 5th Anti-Money Laundering Directive~\cite{5amld}.
Unfortunately,
abuse reporting services can suffer data pollution
due to their crowd-sourced nature.
A fraction of the submitted reports may be useless, including
advertisements, 
trolling,
reports aiming to pollute the database, and
reports not describing abuse.
Some services manually validate the reports
to address data pollution.
Unfortunately, manual validation is costly and does not scale.
Thus, a critical challenge is how to identify valid reports automatically.

Abuse reports are also fundamental for analyzing 
the prevalence and financial impact of different abuse types
(e.g., sextortion~\cite{spamsPaquetClouston},
investment scams~\cite{ponziBartoletti}). 
BBB~\cite{bbb_annual_2023} and FTC~\cite{ftc2023report} 
publish well-known annual reports with statistics on the
received abuse reports,
which prior work often cites in their  
motivation~\cite{bitaab2023beyond,kotzias2023scamdog,kotzias2025ctrl}. 
For example, in 2023, the FTC received 47K abuse reports 
involving cryptocurrency payments,
totaling \$1.4 billion in losses~\cite{ftc2023report}.
Those statistics rely on an accurate classification of the 
abuse reports into abuse types,
for which two approaches exist.
The first approach requires the reporter 
to select the abuse type from a service-specific list of options.
However, reporters 
may not care about accurately classifying the reports, 
may be under stress when reporting abuse, 
may not have enough background to differentiate abuse types, or 
may even be malicious.
An alternative approach is for human analysts to examine the submitted reports, 
inferring the abuse type from the textual description 
where reporters detail the abuse in their own words.
This approach is manual and does not scale well. 
Furthermore, humans can be inconsistent in repetitive tasks. 

This work analyzes the challenge of 
data pollution in cryptocurrency abuse reporting services.
To enable our analysis, we collect cryptocurrency abuse reports 
from two services. 
We obtained 287K abuse reports for 92K Bitcoin addresses 
received by BitcoinAbuse~\cite{bitcoinabuse} over 6 years.
We also collected
\numreportsbbb cryptocurrency abuse reports received by  
Scam Tracker~\cite{scamtracker} over 5 years.

Our work answers the following three research questions: 

\paragraph{(RQ1) What is the extent and impact of pollution in abuse reporting
services?}
We examined three types of pollution: 
spam reports, 
incorrect abuse types reported by users, and 
benign addresses reported as malicious.
We focused on \ba, where there was no filtering for submitted reports.
Then, we checked if the identified issues also affected Scam Tracker,
where reports are manually validated by analysts.
We identify two prevalent types of spam reports in \ba:
advertisements for funds recovery scams and 
reports that aim to discredit cryptocurrency exchanges by claiming they 
are involved in terrorism.
At least 10.6\% of all \ba reports are spam, with the ratio
reaching 75\% of submitted reports prior to the service stopping operation.
In contrast, the manual validation keeps Scam Tracker largely free of spam, 
although we still observe a handful (0.4\%) of missed funds recovery scams.

To measure the accuracy of user-reported abuse types,
we built a ground truth dataset of \numgtfinal abuse report
descriptions labeled with 19 popular abuse types.
\updated{To build this ground truth, we use a two-step approach that 
combines unsupervised clustering to identify large groups of reports with 
similar text and extensive manual labeling.}
We compare this ground truth dataset with the abuse type that 
the reporter selected in the online form, 
among 6 offered options that we call \emph{BA types}.
Reports labeled with three BA types 
(Bitcoin-Tumbler, Darknet-Market, Ransomware)
were almost entirely spam.
Two other BA types (Blackmail-Scam, Other) were too broad. 
The only accurate BA type was Sextortion. 
Even after removing spam reports, users often selected an incorrect 
abuse type, with 22\% of Ransomware-labeled reports
being indeed Sextortion.
In Scam Tracker, analysts classify the reports. 
However, all cryptocurrency abuse reports are placed in the 
same class, preventing fine-grained analysis of abuse types.

Out of the 92,151 Bitcoin addresses reported to \ba, 
\updated{we identify 91 (0.1\%) as benign
using public sources}~\cite{glasschain,arkham,graphsense,wyb}
(most often as belonging to exchanges).
Most (72\%) reports for those benign addresses are spam, 
but we also observe users trying to track the destination of their funds 
and ending up incorrectly reporting an exchange address.
Even if benign addresses are a minority, 
their impact is overwhelming, with those 91 addresses being responsible for 60\% of all received funds. 
While Scam Tracker does not have a specific field for reporting 
cryptocurrency addresses, 
none of the 96 Bitcoin addresses found in the text of reports are 
flagged as benign by the used sources.

Previous works that analyzed 
\ba~\cite{choi2022large,buil2022offending,rosenquist2024dark}
did not filter out spam reports, 
considering that all reported addresses were malicious
(and thus that their deposits corresponded to cybercriminal revenue) and 
that the reporter-selected abuse types were correct. 
Our results cast a shadow over their conclusions.  

\paragraph{(RQ2) Can the identification of valid reports 
and their classification into abuse types be automated?}
We propose a classifier
that leverages a large language model (LLM)
for interpreting the textual description in an abuse report and
assigning it one of 19 abuse types.
Our unsupervised LLM-based classifier does not need training.
It relies on natural language definitions for the abuse types, 
making it easy to extend by adding definitions for new abuse types.
We examine multiple designs for the classifier covering
\updated{the impact of the taxonomy, prompt design, three query approaches, and 
six LLMs}: two open models (llama3, llama3.1) and 
four commercial models (gpt-4, gpt-4o, gpt-4o-mini, gpt-3.5).

We compare our LLM-based classifier with two baselines: 
a naive LLM-based approach and a supervised machine learning classifier.
The best design for the LLM-based classifier 
achieves an F1 score of 0.95 
compared to 0.42 for the naive LLM usage baseline,
highlighting the importance of our design in achieving good accuracy.
In out-of-distribution (i.e., previously unseen) data, it achieves 
a precision of 0.92, a recall of 0.87, and an F1 score of 0.89,
whereas the baseline supervised classifier achieves an F1 score of 0.55.
Furthermore, our LLM-based classifier filters spam reports with 
an F1 score of 0.99.
Finally, we show the ease of extending our classifier with  
a new abuse type.

\paragraph{(RQ3) What is the financial impact of different cryptocurrency 
abuse types?} 
Two approaches exist to estimate the financial impact of 
cryptocurrency abuse types: 
adding losses disclosed by victims in their abuse reports and 
analyzing transactions in the blockchain 
that deposit funds to reported addresses. 
Both approaches require accurate classification of abuse types.
Unfortunately, Scam Tracker does not sub-classify 
cryptocurrency abuse reports, and 
we show that \ba types are untrustworthy.

To answer this question, we first classify
\updated{the 2.3K Scam Tracker reports and a random subset of 5K \ba reports}
using our LLM-based classifier.
Using the classification, we unveil that victims often fall for
investment scams, 
with 61\% of victims of that scam type reporting financial losses, 
four times higher than the next abuse type. 
The median loss is highest for funds recovery scams 
at \$1,700 per report, followed by romance (\$1,500) and 
investment scams (\$1,300).
To estimate revenue from address deposits, we
propose a majority vote to aggregate reports for the same address and
use it to classify 1,930 addresses reported in \ba.
The deposits to those addresses total \$453M and 
confirm that investment scams have the highest financial impact 
(44\% of the revenue).
Despite their low conversion rate, extortions collect nearly \$10M in revenue 
due to their massive email campaigns.
We compare both estimation approaches on 
71 Scam Tracker reports mentioning 96 addresses.
Victim-reported losses heavily underestimate revenue, 
with the estimation from deposit transactions being 29 times higher.

\updated{Our work applies LLMs for interpretive classification of 
cryptocurrency abuse reports, which is part of the broader trend 
towards generative AI in cybersecurity. 
In addition to the text label for each abuse report,
our LLM-based classifier generates a reasoning
that describes why the abuse type was selected,
which we found fundamental for explainability.
It enables the refinement of our abuse type taxonomy and
definitions, required by the LLM-based classifier.
}

\section{Abuse Reporting Services}
\label{sec:overview}

Abuse reporting services receive reports through online forms, 
which include a \emph{description} field for the reporter to 
describe the abuse in natural language.
\updated{Beyond that, we observe the following key differences across services:

\begin{enumerate}

\item Some abuse reporting services, such as 
BitcoinAbuse~\cite{bitcoinabuse}, ChainAbuse~\cite{chainabuse}, and 
BBB's Scam Tracker~\cite{scamtracker}, make the abuse reports public, 
while others, such as FTC's ReportFraud~\cite{reportfraud} and 
ACCC's ScamWatch~\cite{scamwatch}, do not.
We focus on \ba and Scam Tracker because they make abuse reports publicly accessible.
We skipped ChainAbuse (which acquired \ba~\cite{baAcquisition})
because free access to their API is very restrictive,
and they declined our request
for research access to their commercial API.
BitcoinAbuse reports are still available through ChainAbuse.

\item Services may differ in their focus. 
BitcoinAbuse focused on cryptocurrency abuse, accepting 
reports from any geographical location written in different languages. 
ChainAbuse originally focused on cryptocurrency abuse, but 
has since broadened to include other abuse types like phishing.
Scam Tracker covers any abuse type affecting users in the USA and Canada, 
regardless of whether they involve cryptocurrencies.
Thus, reports are in English.

\item Some services post the submitted reports immediately online 
without validation
(e.g., BitcoinAbuse) or after undisclosed filtering (e.g., ChainAbuse).
Others, like Scam Tracker, manually validate the submitted reports
before posting them, determining if they are valid and cleaning up their contents
(e.g., anonymizing personal information).

\item The information requested in the online form varies. 
Scam Tracker allows the reporter to disclose the losses suffered, while BitcoinAbuse did not.
BitcoinAbuse only requested abusers' blockchain addresses. 
In contrast, Scam Tracker does not request blockchain addresses,
which may only appear in the report's text, 
but requests abusers' URLs, email addresses, and phone numbers.

\end{enumerate}
}

As summarized in Table~\ref{tab:datasets}, 
we collect abuse reports from two services: 
\ba and Scam Tracker.

\paragraph{\BA.}
Originally, \ba allowed to download the whole database of abuse reports 
through its API. 
We downloaded a complete database dump on February 1, 2021.
Later, the API changed to only provide the reports of the last 30 days, and 
we have monthly dumps from February 2023 until early June 2023, when 
\ba stopped accepting reports and started redirecting users 
to ChainAbuse~\cite{chainabuse}. 
To fill the collection gap, we built a crawler for the 
\ba webpage, which lists all reported addresses. 
We query the API for each address in the webpage that was not in our dump.

Each report contains the report creation date and 
four fields from the online form: 
the reported Bitcoin address, 
the \emph{abuse type} selected from a drop-down list with 6 options,
an \emph{other-abuse} short string 
optionally used to describe other types of abuse, and 
the description text for freely describing the abuse 
(2,000 characters maximum).
To preserve anonymity, the reports do not identify the reporter. 

\begin{table}
\centering
\scriptsize
\begin{tabular}{lrrrrr}
\hline
\textbf{Dataset} & \textbf{Reports} & \textbf{Desc.} & \textbf{Addr.} & 
\textbf{Start} & \textbf{End} \\
\hline
\ba & \numreports & \numdescriptions & \numaddr & 03/2017 & 06/2023 \\
Scam Tracker & \numreportsbbb & \numdescbbb & 96 & 02/2017 & 05/2024 \\
\hline
\end{tabular}
\caption{Abuse report datasets used.}
\label{tab:datasets}
\end{table}

The dataset comprises \numreports reports for \numaddr Bitcoin addresses,
an average of 3.1 reports per address.
The \numreports reports
have \numdescriptions unique descriptions (by SHA256 of the description text).
Multiple reports may have the same description, 
e.g., due to short descriptions such as ``Sextortion'' or ``Scam''.
We identify the description's language using the 
\emph{langdetect} tool~\cite{langdetect}.
Most descriptions are written in English (80.5\%), followed by 
Russian (1.7\%),
French (1.7\%),
Spanish (1.6\%), and 
German (1.6\%).

\paragraph{Scam Tracker.}
We build a crawler to download abuse reports from the Scam Tracker website.
Scam Tracker groups all reports related to cryptocurrencies 
in the \emph{Cryptocurrency} abuse type. 
In May 2024, we downloaded all \numreportsbbb Cryptocurrency reports.
For each report, we obtain the 
scam ID, report date, description, victim's location, and
dollars lost (optional). 
The \numreportsbbb reports have \numdescbbb descriptions written in English.
The online form does not collect cryptocurrency addresses, 
\updated{but victims may include them in the textual description.
We use the tool \iocsearcher~\cite{iocsearcher}
to extract 96 Bitcoin addresses present in the description field of 71 reports.}

\begin{figure*}[t]
\centering
	\includegraphics[scale=0.8]{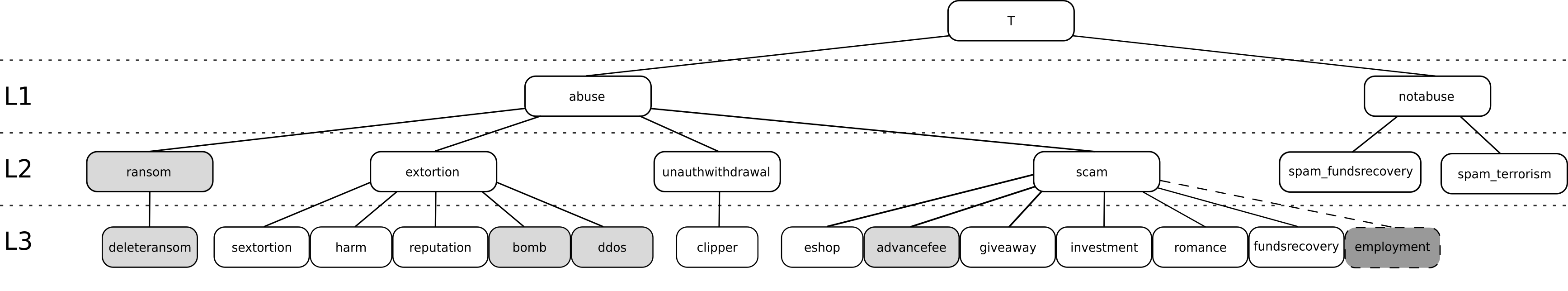}
\caption{Our taxonomy. White nodes come from the initial taxonomy. 
Light gray solid nodes were identified by the clustering.
Dark gray dashed nodes were not originally part of the taxonomy but 
were added during our evaluation.
}
\label{fig:taxonomy}
\end{figure*}

\section{Abuse Type Taxonomy and Ground Truth}
\label{sec:gt-taxonomy}

To answer RQ1 and RQ2,
we build a taxonomy of prevalent abuse types and 
ground truth (GT) datasets of descriptions annotated with their 
abuse type in the taxonomy.
Section~\ref{sec:taxonomy} describes the design of our taxonomy, and
Section~\ref{sec:construction} details the process for 
generating the taxonomy and ground truth.

\subsection{Taxonomy}
\label{sec:taxonomy}

Prior work has presented cryptocurrency abuse 
taxonomies~\cite{vasek2015there,badawi2020cryptocurrencies,bartoletti2021cryptocurrency}. 
Our taxonomy builds on those, but it is more complete and 
specifically designed hierarchically to better accommodate possible 
incompleteness and facilitate extensions, as explained next.
Figure~\ref{fig:taxonomy} shows our tree-structured abuse taxonomy,
where nodes are abuse types, and 
edges capture parent-child relationships,
e.g., \ttag{investment} is a subtype of \ttag{scam}. 
The lower a node, the more specific abuse it captures.
\updated{Next, we detail each of the levels underneath the root node ($\top$).}

\paragraph{First level (L1).}
It captures whether the report is valid (\ttag{abuse})
or useless (\ttag{notabuse}).
Valid abuse reports describe abuse suffered or observed by the 
reporter, regardless of whether the abuse was successful
(i.e., lost funds).
In contrast, \ttag{notabuse} reports may 
advertise a service,
try to pollute the reporting database (e.g., trolling), or 
offer no usable information 
(e.g., provide a personal opinion of the reporter).
The two L1 nodes cover all reports, 
i.e., a report is either an abuse report or not.

\paragraph{Second level (L2).}
It captures the four main abuse types identified
and two prevalent types of spam reports.
A \ttag{scam} is an attempt to defraud by gaining 
the victim's confidence.
The term \ttag{scam} is sometimes used 
to refer to any abuse type, i.e., as a synonym of \ttag{abuse}. 
However, in this work, we use \ttag{scam} to refer only to attacks that abuse 
the user's confidence.
In contrast, \ttag{extortion} captures coercion, 
i.e., the victim is threatened with some future harm
if a payment to the extortionist is not made.
An extortionist does not try to gain the victim's confidence as a 
scammer does but directly threatens the victim.
Some attacks demand a \ttag{ransom} payment from the victim, 
e.g., to recover access to encrypted or deleted data. 
The difference with \ttag{extortion} is that there is no threat of future harm.
Instead, in \ttag{ransom}, the damage has already happened, 
i.e., the data has already been encrypted or deleted.
Finally, \ttag{unauthwithdrawal} captures thefts where
funds belonging to a victim are transferred 
without authorization or consent,
e.g., by stealing addresses' private keys.
This level also includes two prevalent types of \ttag{notabuse} spam reports. 
The first is reports advertising funds recovery services
(\ttag{spam\_fundsrecovery})
where reporters state that they lost funds to some scam and
then used a funds recovery service to get the funds back.
The report provides a contact (i.e., email, URL, phone) 
for the funds recovery service.
Given the difficulty of recovering Bitcoin funds,
even for law enforcement,
these services are with all probability scams.
The other are reports claiming that the reported addresses
are involved in terrorism (\ttag{spam\_terrorism}),
but those addresses belong to exchanges (e.g., Binance, HTX).

\paragraph{Third level (L3).}
It captures prevalent subtypes of the main abuse types.
We identify 5 subtypes of \ttag{extortion}.
In \ttag{sextortion}, the extortionist threatens to 
release the victim's data of sexual character, most often a video. 
But, the extortionist may also threaten
to damage the victim's \ttag{reputation} 
or personal integrity (\ttag{harm}), 
make some resources of the victim unavailable (\ttag{ddos}), or 
damage other persons known to the victim by planting a \ttag{bomb}. 
Note that the \ttag{ddos} node is underneath \ttag{extortion} 
because in the reports we observe the attacker 
threatens to launch a DDoS attack in the future if a payment is not made.
If the attacker had instead already launched a DDoS attack and requested 
a payment to stop it, we could add another node under \ttag{ransom}.
We identify 7 prevalent \ttag{scam} subtypes.
The attacker may try to convince the victim to participate 
in a supposedly highly profitable \ttag{investment}
or in a \ttag{giveaway} that promises victims 
they will receive back twice (or more)
the amount they send to the scammers.
Scammers may set up a fake online \ttag{eshop} that does not deliver 
the goods bought by the victim, 
request a fee in advance for a service 
that will not be delivered (\ttag{advancefee}), 
try to establish a friendship or loving 
relation to later request money from the victim (\ttag{romance}), 
offer shady jobs that are not paid (\ttag{employment}), or
promote useless funds recovery services (\ttag{fundsrecovery}).
Note that \ttag{fundsrecovery} corresponds to valid reports of victims 
scammed by funds recovery services, while 
\ttag{spam\_fundsrecovery} are advertisements posted by scammers.
For \ttag{unauthwithdrawal}, we only observe one prevalent subtype
corresponding to the victim's device becoming infected with a \ttag{clipper} 
malware that replaces addresses copied by the user (i.e., in the clipboard)
with attacker-controlled addresses.
If the victim uses the (replaced) address as the destination 
in a transaction, the attacker will receive the funds.
For \ttag{ransom}, we only observe one prevalent subtype 
that deletes data,
typically from a database with no or weak authentication,
and requests a ransom to return the deleted data 
(\ttag{deleteransom})~\cite{serverransom}. 

\paragraph{Granularity.}
One may wonder whether all nodes in L3 are really needed, 
e.g., whether it is needed to differentiate \ttag{bomb} and \ttag{ddos} 
from other extortions.
Our approach has been to add all prevalent abuse types that we have observed
because this helps to better characterize social engineering tactics 
used by attackers.
However, the taxonomy's hierarchical design allows to easily remove 
fine-grained abuse types. 
For example, if \ttag{bomb} and \ttag{ddos} 
are considered unnecessary and removed from the 
classifier in Section~\ref{sec:classification}, 
those reports will then be labeled using the 
more general \ttag{extortion} parent.

\paragraph{Completeness.}
Any abuse taxonomy is also necessarily incomplete
as cybercriminals frequently introduce new abuse types and variations.
Our hierarchical taxonomy is designed to handle such missing abuse types.
For example, any scam report that does not match the 7 L3 scam subtypes 
can be labeled using the L2 \ttag{scam} type, 
indicating that it captures another type of scam.
Examining the reports assigned an L2 type (or the L1 \ttag{abuse} type) may 
reveal new prevalent abuse types that should be added to the taxonomy.
For example, \ttag{employment} scams
were not originally part of the taxonomy but were discovered by examining 
the reports classified as \ttag{scam} in Section~\ref{sec:updates}.

\subsection{Ground Truth Construction}
\label{sec:construction}

Since we cannot manually examine the \numdescriptions 
\ba descriptions,
we design a two-step approach to build 
our taxonomy in Figure~\ref{fig:taxonomy} and 
the GT datasets in Table~\ref{tab:gt}.
First, we manually analyze \numgtinitial descriptions
to produce an initial GT (gt\_initial) 
with its corresponding initial taxonomy.
This step identifes the 16 abuse types in white in Figure~\ref{fig:taxonomy}.
Then, we cluster the 224K descriptions
by the similarity of their text and
manually label a fraction of the clusters.
The cluster analysis identifies 5 additional prevalent abuse types
not in gt\_initial,
which we add to the taxonomy (light gray nodes in Figure~\ref{fig:taxonomy}).
It also produces a larger GT (gt\_final) 
with \numgtfinal descriptions.

\begin{table}
\centering
\scriptsize
\begin{tabular}{lrrr}
\hline
\textbf{Type} & \textbf{gt\_initial} & \textbf{gt\_final} & \textbf{gt\_bbb} \\
\hline
	sextortion & 44 (25.1\%) & 16,112 (82.8\%) & 23 (11.5\%) \\
	extortion & 28 (16.0\%) & 1,417 (7.3\%) & 2 (1.0\%) \\
	giveaway & 6 ( 3.4\%) & 740 (3.8\%) & 6 (3.0\%) \\
		spam\_fundsrecovery & 1 ( 0.6\%) & 174 (0.9\%) & - \\
	harm & 1 ( 0.6\%) & 170 (0.9\%) & - \\
	deleteransom & - & 162 (0.8\%) & 1 (0.5\%) \\
	investment & 11 ( 6.3\%) & 143 (0.7\%) & 102 (51.0\%) \\
	advancefee & - & 74 (0.4\%) & 8 (4.0\%) \\
	notabuse & 27 (15.4\%) & 73 (0.4\%) & 8 (4.0\%) \\
	spam\_terrorism & 3 ( 1.7\%) & 64 (0.3\%) & - \\
	ddos & - & 62 (0.3\%) & - \\
	scam & 31 (17.7\%) & 60 (0.3\%) & 48 (24.0\%) \\
	reputation & 1 ( 0.6\%) & 54 (0.3\%) & - \\
	bomb & - & 52 (0.3\%) & - \\
	romance & 2 ( 1.1\%) & 34 (0.2\%) & - \\
	clipper & 1 ( 0.6\%) & 33 (0.2\%) & - \\
	unauthwithdrawal & 7 ( 4.0\%) & 7 (0.0\%) & 2 (1.0\%) \\
	eshop & 6 ( 3.4\%) & 6 (0.0\%) & - \\
	abuse & 5 ( 2.8\%) & 5 (0.0\%) & - \\
	fundsrecovery & 1 ( 0.6\%) & 1 (0.0\%) & - \\
  ransom & - & - & - \\
	\hline
	All & \numgtinitial (100\%) & \numgtfinal (100\%) & \numbbb (100\%) \\
\hline
\end{tabular}
\caption{Descriptions for each abuse type in GT datasets.}
\label{tab:gt}
\end{table}

\paragraph{Initial ground truth and taxonomy.}
We built an initial taxonomy
by manually labeling \numgtinitial descriptions. 
To select these descriptions, we split the 
287K reports in the \ba dataset per year of the report 
and the six BA types,
producing 35 buckets:
one per year (from 2018 to 2023)
and abuse type
(except for Sextortion in 2018 since it was created in 2019).
Then, we randomly selected 5 unique descriptions from each bucket, 
avoiding duplicates.
Two analysts manually classified each of the \numgtinitial descriptions
by reading their text and freely assigning labels to the descriptions 
based on their experience.
After completing the first labeling round, the analysts discussed the 
selected labels, normalized their names, and merged duplicates,
coming up with the 16 abuse types in white in  Figure~\ref{fig:taxonomy}.
Thus, despite the small number of descriptions in gt\_initial,
this analysis identifies 16 out of the 22 entries in the final taxonomy.
Then, the analysts performed a second joint labeling round 
to assign one of the 16 identified abuse types to the
\numgtinitial descriptions.

\paragraph{Clustering.}
To discover additional abuse types not present in gt\_initial
and to produce a larger ground truth,
we cluster the 224K descriptions by the syntactical similarity of their text.
We select the \emph{all-mpnet-base-v2} pre-trained embedding from the 
\updated{Hugging Face repository~\cite{all-mpnet-base-v2}}, 
a general-purpose model that achieves the highest average performance 
for sentence similarity 
\updated{in the Sentence Transformers evaluation}~\cite{huggingface-models}.

To cluster the embeddings, we use
\textit{hierarchical agglomerative density-based clustering}
\updated{(HDBSCAN)~\cite{hdbscan}}
because it can find clusters of arbitrary shape 
and does not require knowing the number of clusters in advance.
We use Euclidean distance.
\updated{We set the minimum core point neighborhood size
to a low value of 2 to allow the discovery of less dense clusters.
We set the minimum cluster size to 5 so that the cluster captures 
at least a handful of reports.} 
The clustering produces 5,265 clusters containing one-third of the descriptions.
The largest cluster has 3,814 descriptions.
The remaining two-thirds (66.3\%) of descriptions
end in singleton clusters.

One of the analysts examined the largest 150 clusters.
Most clusters (95) captured \ttag{sextortion} email campaigns,
followed by 9 \ttag{giveaway} clusters capturing reports 
by security vendors or analysts.
This process identifies four new abuse types:
\ttag{deleteransom}, \ttag{advancefee}, \ttag{ddos}, and \ttag{bomb}.
We additionally introduce \ttag{ransom} as the parent of \ttag{deleteransom},
to capture ransomware reports.
To check the smaller clusters,
we produce a word cloud~\cite{wordcloud} for each of the 5,265 clusters.
The analyst visually examined the word clouds,
selecting 30 clusters to analyze because they contain words that 
differ from those in the top 150 clusters. 
However, all 30 clusters corresponded to abuse types already 
in the taxonomy.

We manually examined 175 singletons selected 
using the same approach as gt\_initial
(i.e., 35 buckets, 5 descriptions from each).
Of the 175 descriptions,
99 (56\%) belonged to a fine-grained L3 abuse type, and
26 (15\%) were \ttag{notabuse}.
Among the 50 (29\%) descriptions assigned an L1 or L2 abuse type,
there were no new abuse types.
Thus, the examined singletons correspond to syntactic variations
of abuse types already in the taxonomy.

\paragraph{Final taxonomy and ground truth.}
Figure~\ref{fig:taxonomy} shows the final taxonomy,
where the light gray nodes correspond to the 5 abuse types introduced by the
cluster analysis.
Table~\ref{tab:gt} describes the final ground truth (gt\_final), 
which contains all gt\_initial descriptions, 
plus those in the clusters manually labeled by the analyst.
In total, gt\_final contains \numgtfinal descriptions.
The top abuse type is \ttag{sextortion} with 16,112 (82.8\%) descriptions,
followed by \ttag{extortion} (7.3\%),
\ttag{giveaway} (3.8\%), and
\ttag{spam\_fundsrecovery} (0.9\%).

\paragraph{Scam Tracker ground truth.}
We apply the same process for generating gt\_initial to 200 
Scam Tracker descriptions (gt\_bbb).
The abuse type distribution in gt\_bbb differs from that in gt\_final 
with half of the abuse reports (51\%) being for \ttag{investment} scams
followed by general scams (24\%), and \ttag{sextortion} (11.5\%).
\updated{The fact that gt\_bbb contains a larger fraction of scam reports, 
compared to the dominant sextortion reports in gt\_final,}
makes gt\_bbb ideal 
to validate our classifiers, making sure they are not overfit to the 
abuse reports in gt\_final.

\section{Data Pollution in Abuse Reporting Services}
\label{sec:pollution}

This section answers RQ1 by analyzing 
data pollution in
abuse reporting services. 
We focus on \BA because it did not validate the submitted reports.
Additionally, we examine which of the identified issues affect Scam Tracker.

\subsection{Spam Reports}
\label{sec:spam}

Section~\ref{sec:construction} identified two prevalent spam report types
in \ba:
advertisements for funds recovery services (\ttag{spam\_fundsrecovery}) and 
fake terrorism reports (\ttag{spam\_terrorism}).
In this section, we measure the volume of those spam reports 
across the whole \ba dataset. 

\updated{Spammers may use the same description text in 
hundreds or even thousands of abuse reports.}
For example, the 174 \ttag{spam\_fundsrecovery} descriptions
in gt\_final (compared to 16,112 for \ttag{sextortion})
appear in 21,070 reports
(compared to 17,837 reports for \ttag{sextortion}),
an average of 121 reports per description
(compared to an average of 1.1 for \ttag{sextortion}).

To identify \ttag{spam\_fundsrecovery} reports across all the \ba data, 
\updated{we build a blocklist with indicators appearing in the 174 
\ttag{spam\_fundsrecovery} descriptions in gt\_final. 
These indicators are used by scammers to tell potential victims how to 
contact the funds recovery service}.
The blocklist contains 27 domains, 17 email addresses, 12 phone numbers, 
10 URLs, and one Telegram handle.
Then, we scan the 287K \ba reports using the blocklist. 
We conservatively match domains so that if the domain \url{summitrecoup.com} 
is in the blocklist, we match any reports containing the 
string \emph{summitrecoup}.
This process identifies 208 (0.1\%) \ttag{spam\_fundsrecovery}
descriptions used in 22,180 (9.9\%) reports.
To identify \ttag{spam\_terrorism} reports across all the \ba data, 
we manually examine the 119 descriptions not in gt\_final that contain 
the string \emph{terror}.
This process identifies 79 (0.03\%) \ttag{spam\_terrorism}
descriptions used in 1,500 (0.7\%) reports.

\begin{figure}
\centering
\includegraphics[width=\columnwidth]{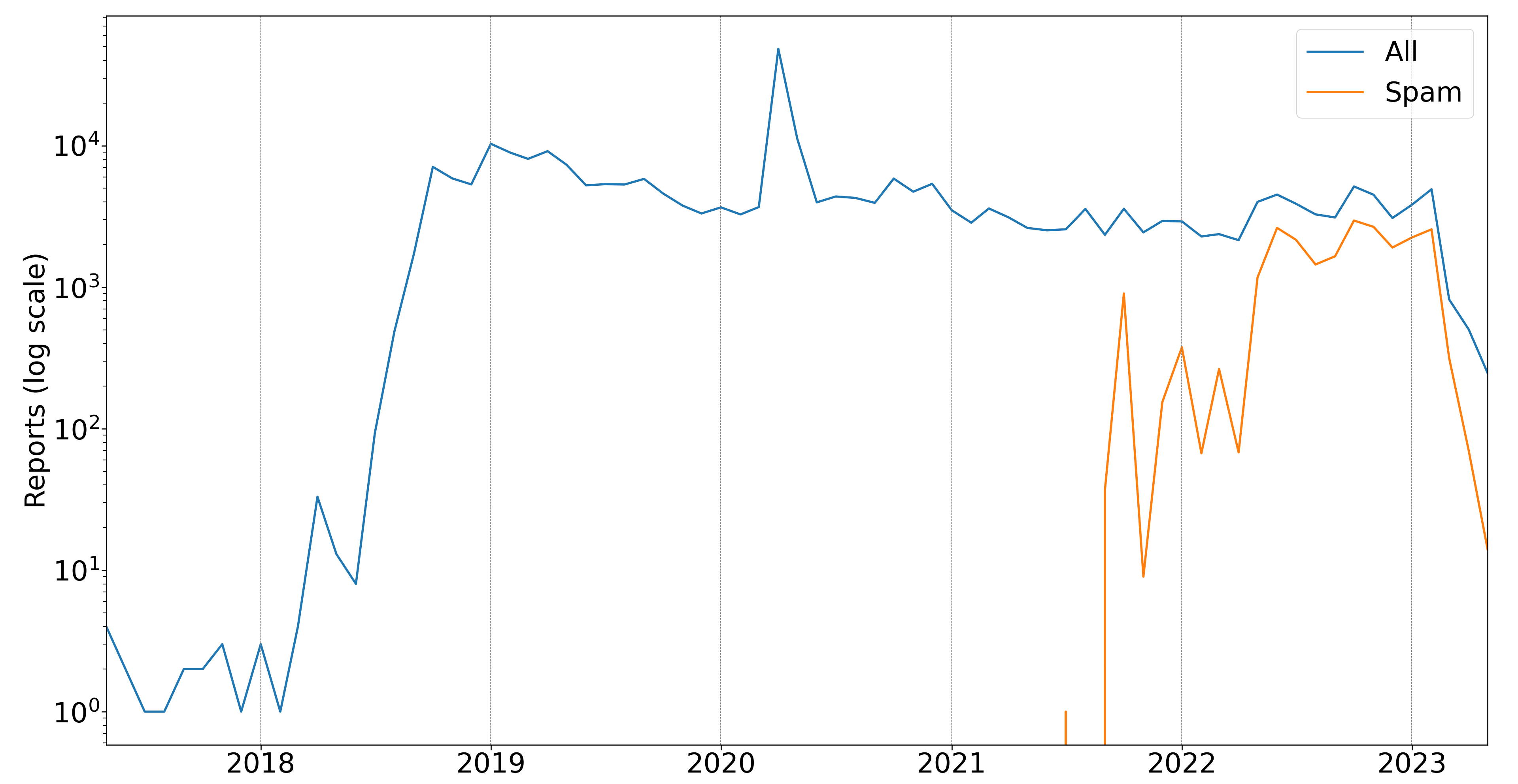}
\caption{Volume of \ba submitted reports and spam reports 
(log-scale) per month.
Spam reports are those labeled \ttag{spam\_fundsrecovery} and 
\ttag{spam\_terrorism}.} 
\label{fig:spam}
\end{figure}

Figure~\ref{fig:spam} shows the volume of 
spam reports (\ttag{spam\_fundsrecovery}, \ttag{spam\_terrorism}) 
and all reports over time.
The volume of spam identified is a lower bound, 
as smaller spam types may also exist.
The two prevalent spam types first appeared in February 2021 and 
dominated all reports since mid-2022, 
reaching 75\% of all submitted reports between June and December 2022. 
Despite the quickly rising spam, 
the total volume of reports remained quite stable during 2021-2022, 
indicating that as spam increased the number of 
valid (non-spam) reports decreased, likely because users started to find 
the spammed service less useful.
Submitted reports declined since February 2023 as BitcoinAbuse started 
redirecting users to ChainAbuse, 
eventually stopping the reception of new reports in June 2023.

Our results show that,
when no filtering is applied,
spam will eventually dominate the submitted reports.
In contrast, the manually validated Scam Tracker reports are largely free of 
spam, 
as we only identify 9 \ttag{spam\_fundsrecovery} reports across the entire 
Scam Tracker dataset, as detailed in Section~\ref{sec:spamdetection}.

\subsection{Accuracy of User-Reported Abuse Types}
\label{sec:batypes}

Next, we evaluate the accuracy of user-reported BA types
by comparing them with the taxonomy abuse types in gt\_final.
The evaluation is performed on reports because the same description 
may correspond to multiple reports with different BA types.
Thus, we first propagate the taxonomy types from 
the \numgtfinal descriptions in gt\_final to 43,693 reports
(15.2\% of all \numreports reports)
where those descriptions appear.

\begin{figure*}
\centering
\includegraphics[width=\textwidth]{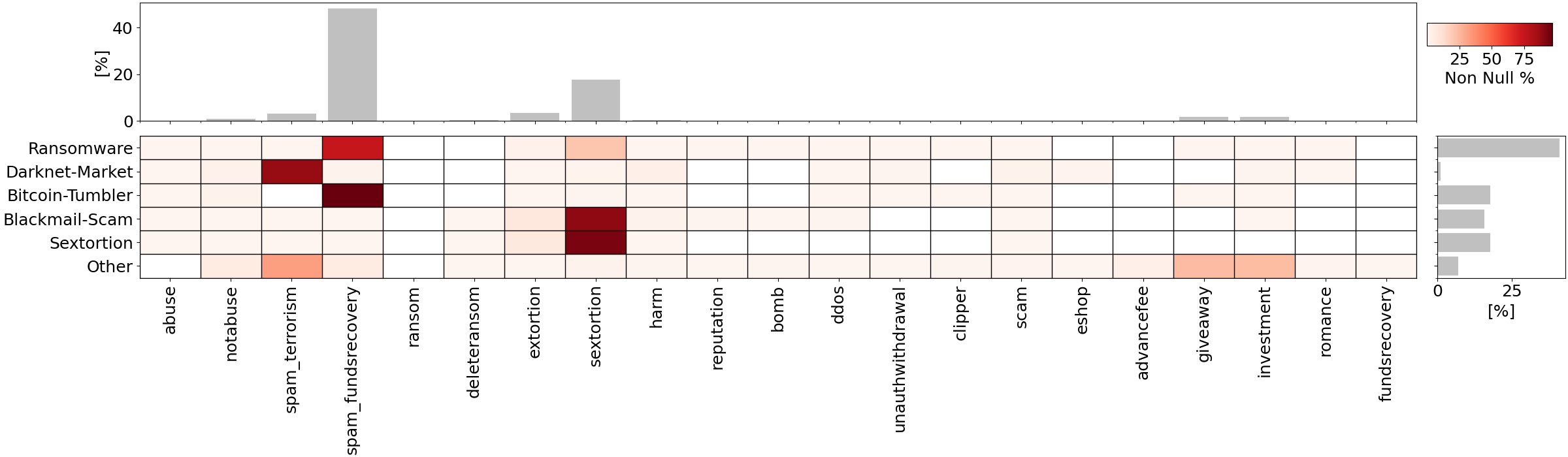}
\caption{Heat-map capturing the fraction of the 43K evaluation reports 
assigned the BA type on the y-axis and tagged 
with the taxonomy abuse type on the x-axis.
White means the ratio is zero,
light orange is a small ratio, and
dark red is a high ratio.
}
\label{fig:batypes-heatmap}
\end{figure*}

Figure~\ref{fig:batypes-heatmap} shows a heat map 
where the heat captures the fraction of the 43K reports assigned
the BA type on the y-axis and tagged with the taxonomy 
abuse type on the x-axis. 
White means the ratio is zero, 
light orange means a small ratio, and 
dark red a high ratio.
The bars on top capture the fraction of the 43K reports
assigned to each taxonomy abuse type.
The bars on the right capture the fraction of the 43K reports 
assigned to each BA type.

\paragraph{Spam impact.}
Four of the six BA types are dominated by spam reports, namely 
Bitcoin-Tumbler (98\%),
Darknet-Market (91\%), 
Ransomware (75\%), and
Other (43\%).
For these four BA types, any conclusions drawn
without filtering out the spam reports
would be heavily biased toward the spam reports
and thus would be incorrect.

\paragraph{Sextortion.}
Of the 7,766 reports with the Sextortion BA type,  
97.4\% are indeed \ttag{sextortion}, and 2.5\% are \ttag{extortion}.
Among the 2.5\% \ttag{extortion} reports, some are likely 
in reality \ttag{sextortion} where the reporter only copied a small part of the 
received email in the description 
that did not mention the sexual character of the extortion, 
making our analysts label it \ttag{extortion}.
Thus, the Sextortion BA type correctly captures \ttag{sextortion} abuse. 
On the other hand, the most common misclassification 
(after excluding spam reports) is
22\% Ransomware reports that are instead \ttag{sextortion}.
A possible reason is that reporters who did not know, or did not care, 
about the abuse type just selected the first option
in the drop-down list (Ransomware).
We conclude that reporters could correctly identify \ttag{sextortion} abuse.
However, most reporters do not care about selecting the 
right abuse type, which indicates that this task needs to be automated
or performed by human analysts. 

\paragraph{Ransomware.}
gt\_final has no \ttag{ransom} reports, 
but there are 196 \ttag{deleteransom} reports. 
Of those, 173 (88.3\%) are correctly assigned the Ransomware BA type.
Unfortunately, \ttag{deleteransom} is only 1\% of Ransomware reports. 
Thus, the Ransomware BA type is untrustworthy,
even after removing spam reports.
We conclude that \ba was not commonly used to report ransomware, 
but when it was (e.g., \ttag{deleteransom}), users identified accurately
this abuse type.

\paragraph{Blackmail-Scam.}
This BA type was too broad, covering all taxonomy entries underneath
\ttag{extortion} and \ttag{scam}.
Of the 6,830 Blackmail-Scam reports,
92\% are \ttag{sextortion}.
Of these, 19\% correspond to reports before the introduction of
the Sextortion BA type in February 2019,
when Blackmail-Scam was the best option for \ttag{sextortion}.
In the rest, the reporter may have considered that Blackmail-Scam
fit the bill and perhaps did not even realize there was another more
specific Sextortion entry further down the list.
We conclude that broad abuse types should be avoided 
as they may cause errors due to overlapping more specific abuse types.

\paragraph{Bitcoin-Tumbler and Darknet-Market.}
All reports in these BA types (even after excluding spam reports) were
incorrectly classified since 
gt\_final does not have abuse types equivalent to 
Bitcoin-Tumbler and Darknet-Market. 
This makes sense as users of mixers and darknet markets may not be inclined
to report the services they use.

\paragraph{Other.}
Among the 3,051 Other reports,
43.1\% are spam reports,
24.0\% are \ttag{giveaway},
23.6\% are \ttag{investment},
and 9\% are scattered across 14 abuse types.
The widespread presence of \ttag{giveaway} and \ttag{investment} scams
may indicate that many reporters know how to identify those abuse types.
However, since the platform did not offer such types to its users, 
they opted for the generic type Other.
We conclude that it is hard to predict what abuse types victims will report
and, thus, which types the online form should offer.

\paragraph{Impact on previous works.}
Previous works analyzing \ba reports rely on the 
BA types~\cite{choi2022large,buil2022offending,rosenquist2024dark}.
Unfortunately, we have shown that user-reported BA types are most 
often incorrect.
Conclusions drawn from the Bitcoin-Tumbler, Darknet-Market, and Ransomware BA 
types will be incorrect as those BA types are dominated by spam, with 
the remaining reports being mostly misclassifications. 
On the other hand, conclusions drawn from Sextortion should 
be trustworthy as most reporters accurately use that BA type for 
\ttag{sextortion} abuse.
Blackmail-Scam and Other are too broad.
Blackmail-Scam is dominated by \ttag{sextortion}, so conclusions drawn 
on this BA type should match those of Sextortion.

Scam Tracker places all cryptocurrency abuse reports under the same category.
Thus, we cannot evaluate the accuracy of the abuse type selection by analysts. 

\subsection{Benign Addresses Reported}
\label{sec:benign}

\begin{figure}
\centering
\includegraphics[width=\columnwidth]{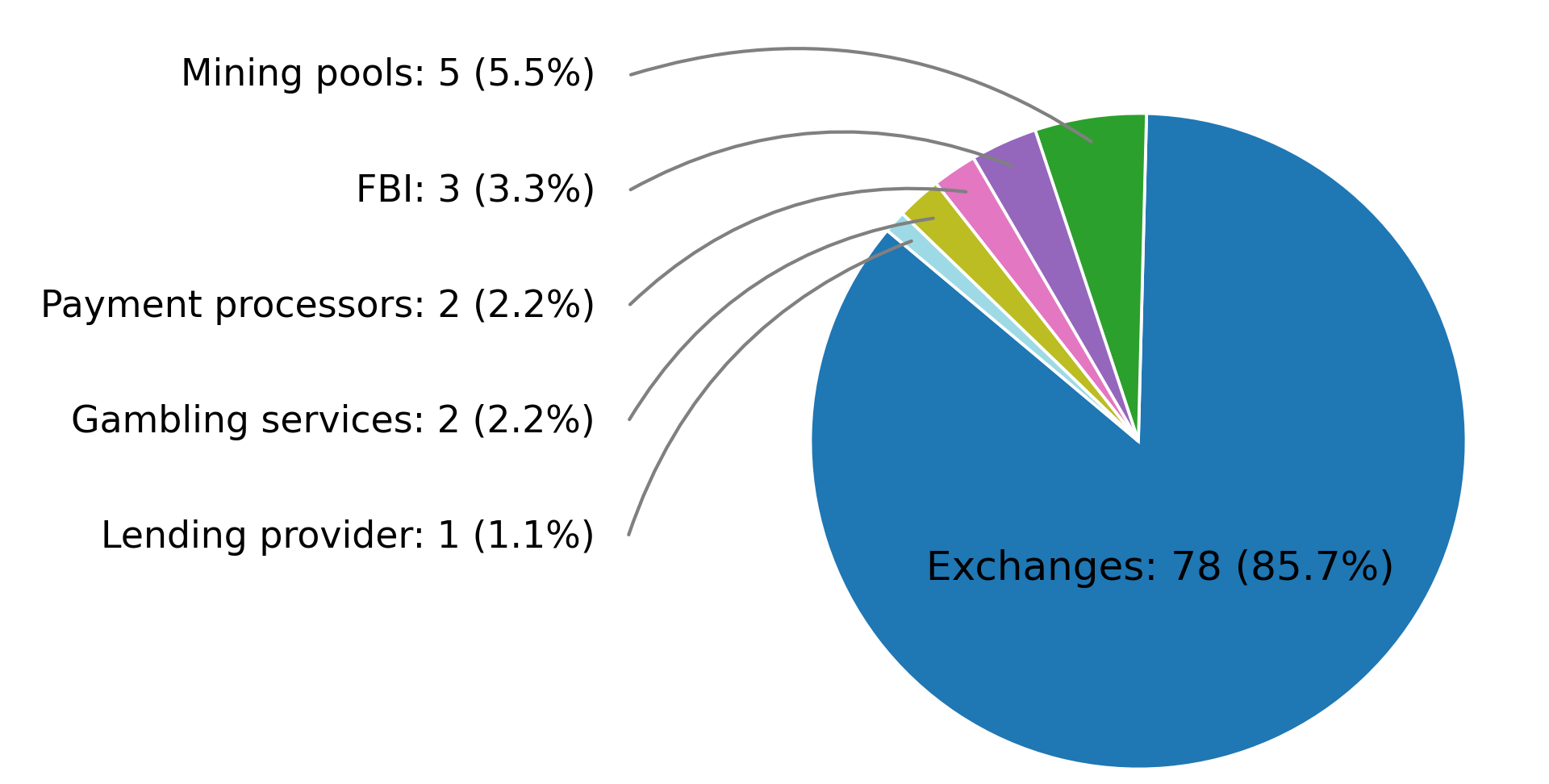}
\caption{\updated{Split by owner category of the 91 benign addresses identified in \ba reports.}} 
\label{fig:benign}
\end{figure}

\BA indexed reports by the reported Bitcoin address.
Next, we check whether users may have reported addresses known to be benign, 
e.g., belonging to services like exchanges. 
For this, we leverage four public address-tagging 
sources~\cite{glasschain,arkham,graphsense,wyb}.
Of the \numaddr reported Bitcoin addresses, 
91 (0.1\%) are tagged in those sources as internal service addresses or 
belonging to benign organizations.
These are mainly hot wallets used for the daily operations of services
and cold wallets used for long-term storage of the service reserves.
\updated{As illustrated in Figure~\ref{fig:benign}, 
of the 91 addresses, 
78 belong to exchanges, five to mining pools,
three to the FBI,
two to payment processors, 
two to gambling services, and 
one to a lending provider.}

\updated{These 91 addresses have 780 reports (0.3\% of the 287K \ba reports).}
Of those, 436 (56\%) are \ttag{spam\_terrorism}
claiming that an exchange is involved in terrorism and 
reporting the hot wallet of the exchange.
Another 123 (16\%) are \ttag{spam\_fundsrecovery}
posted on frequently reported addresses
where they are more likely to be viewed by victims,
i.e., the reported address is unrelated to the funds recovery service.
The other 28\% of reports have two main reasons. 
First, victims may try to track down where their stolen funds are going and 
end up reporting 
an exchange's hot wallet
that received the money after the attackers cashed out.
Victims should be educated to only
report addresses they directly sent funds to (i.e., payment addresses),
without trying to track down the funds,
since information in the blockchain is immutable and
tracking can be done later by experienced analysts.
In other cases, the reporter may want to pollute an exchange's reputation 
similar to \ttag{spam\_terrorism},
e.g., a report asks for a donation to a cold wallet of the 
HTX exchange.

\updated{While only 91 (0.1\% of all reported addresses) are benign,
their impact on estimating cybercrime-related financial losses is huge 
because they may receive large amounts of bitcoins unrelated to cybercrime.
In Section~\ref{sec:estimation}, we show that
the estimation of total losses suffered by \ba reporters
is reduced by 60\% when removing those 91 addresses.}

The Scam Tracker form does not have a field to report 
blockchain addresses. 
We extracted 96 Bitcoin addresses from the descriptions and 
none of those are benign. 

\section{Abuse Report Classification}
\label{sec:classification}

This section answers RQ2 by analyzing different designs for a classifier
to automate identifying valid reports and their classification
into fine-grained abuse types.

\updated{
We approach this as a multi-class classification problem.
Given a report's description,
i.e., the victim's textual description of the abuse,
our classifier returns 
the most specific (i.e., lowest) 
applicable abuse type in our taxonomy for the description.
No service-specific metadata is used
for the classification.
Thus, any abuse report with a textual description can be processed 
regardless of the service the report originates from.
If multiple reports contain the same description, 
the description is classified once and the abuse type 
is propagated to all reports.
If a report has traits of two abuse types 
(e.g., romance and investment~\cite{cross2023romance}), 
our classifier selects only one. The alternative would be using multiple Boolean classifiers, 
one per abuse type.
But, this is problematic because
a report should not be abuse and notabuse.
Furthermore, double-counting a report in two abuse types is problematic 
when computing financial impact.

We investigate two types of classifiers:
an unsupervised classifier that leverages an LLM 
(detailed in Section~\ref{sec:llm}) and 
a machine learning (ML) supervised classifier 
(described in Section~\ref{sec:supervised}).
LLMs are well-suited for our classification task
given their ability to interpret natural language text.
They can capture semantic nuances and
implicit meanings within victim narratives, which are often informal,
varied in expression, and emotionally charged.
On the other hand, we do not want to use an LLM if other alternatives 
exist that can deliver similar accuracy at a fraction of the cost. 
For this reason, we also evaluate a supervised classifier, 
which has advantages in terms of 
better scalability, lower environmental footprint, and lower costs.
We compare the accuracy of both types of classifiers and their 
ability to generalize to abuse reports that differ from those 
previously observed,
which is critical given that abuse reports may describe similar
abuse types in vastly different ways.
We also examine other properties such as the handling of underrepresented
classes, ease of updating, cost, scalability, environmental footprint, and 
explainability.
}

\begin{figure}
\centering
\includegraphics[width=\columnwidth]{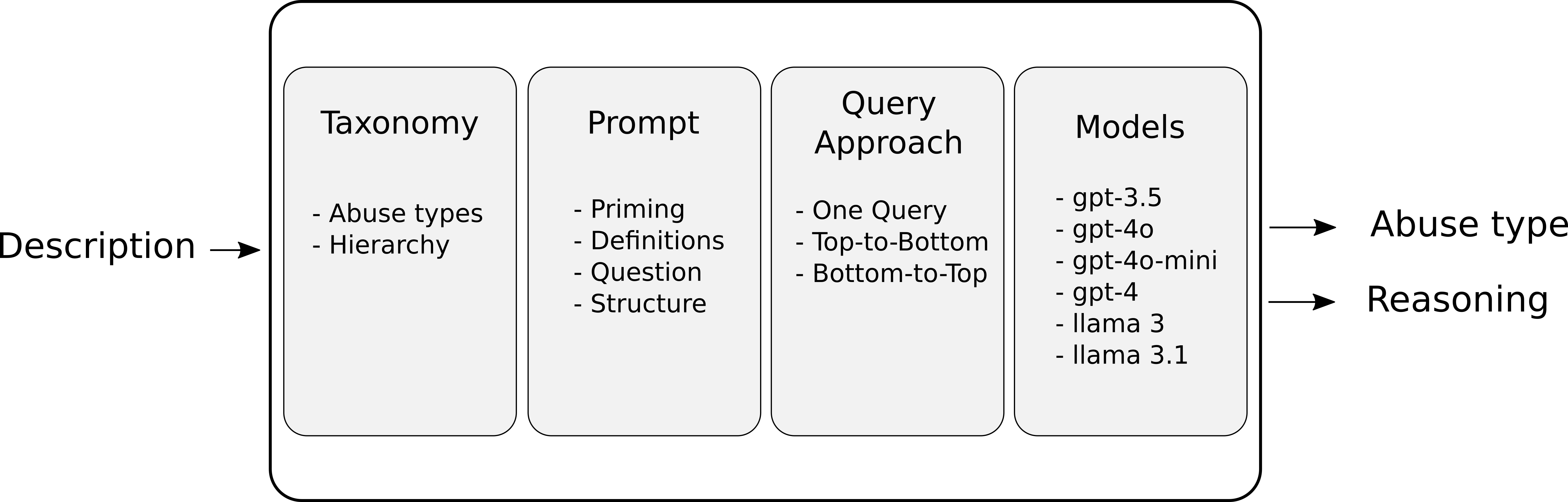}
\caption{\updated{LLM classifier design.}}
\label{fig:design}
\end{figure}

\subsection{\updated{LLM-Based Classifier Design}}
\label{sec:llm}

\updated{Using an LLM to classify an abuse report reduces to 
preparing a prompt that includes the description's text and requests the 
LLM to generate a label for the best-fitting abuse type. 
However, simply providing the description text and asking the LLM to select an 
abuse type in our taxonomy works poorly. 
To illustrate this, we define an \emph{llm-baseline} that uses the 
prompt in Figure~\ref{fig:query2} in the Appendix to do exactly that.
As we show in Section~\ref{sec:eval}, this baseline works poorly. 
We identify two main reasons for its poor accuracy. 
First, the LLM's pre-existing knowledge may contain conflicting definitions of 
the abuse types in our taxonomy.
For example, \ttag{scam} is sometimes used as a synonym for \ttag{abuse}
rather than capturing only attacks that exploit users' confidence.
Conflicting definitions introduce noise, making 
the selected abuse type not match the expected one. 
Second, abuse types are not disjoint.
For example, scams are a subset of all abuses,
and investment scams are a subset of scams.
Thus, the LLM may pick any of the overlapping abuse types. 
Even when instructed to pick the most specific (i.e., lower) abuse type, 
we observe that the overlap often confuses the model.

The baseline's low accuracy
highlights the fundamental need for a careful design
to build an accurate LLM-based classifier.
As illustrated in Figure~\ref{fig:design}, 
we investigate four key design aspects:
the taxonomy (already detailed in Section~\ref{sec:taxonomy}),
the prompt design,
the query approach (i.e., sequence of queries to classify a description), and
the LLM queried.

\paragraph{Prompt design.}
We design the prompt with the primary goal of addressing 
the challenge of conflicting definitions in the LLM pre-trained knowledge. 
In addition, the prompt should respect a few constraints:
it should be blockchain-agnostic, 
so that it can be used to classify reports regardless of the abused blockchain, 
and it should be model-agnostic, 
so that it can be used with multiple LLMs. 
Given the quick release of new models with increased reasoning capabilities, 
optimizing the prompt for a specific model is not currently an efficient 
use of resources.

To address the problem of conflicting definitions 
in the LLM pre-trained knowledge, 
we include our own definitions
for the abuse types of the taxonomy in the prompt.
Then, the prompt requests the LLM to select the output abuse type
based on the provided definitions, 
rather than its pre-trained (possibly conflicting) knowledge.
This definition-based prompting approach has several advantages.
First, it provides the LLM with the scope and context for each abuse type, 
avoiding conflicting definitions it may have learned during training.
Second, it enables identifying under-represented abuse types 
in our ground truth, such as \ttag{fundsrecovery} 
with only one description and \ttag{ransom} with no descriptions at all.
Even if few or no examples are available in the GT, 
we can still write a definition that 
captures the known characteristics of those abuse types and let the classifier
identify matching unlabeled descriptions.
In contrast, a supervised classifier could not identify descriptions of 
such under-represented abuse types. 
Third, it allows for easy updates to the classifier. 
Once a new abuse type is added to the taxonomy, 
we only need to provide a definition for it. 
There is no need for re-training as the classifier is unsupervised.
Fourth, it forces us to specify what each abuse type means
and how it differs from other abuse types, helping us refine our taxonomy, 
e.g., when deciding whether two abuse types should be merged
due to overlaps in their definitions.
At the end, each taxonomy entry is associated with a definition.
Users can leverage such definitions
to understand the scope and characteristics of the abuse types.
}

While constructing the abuse taxonomy,
the analysts generated an initial list of abuse type definitions.
Those initial definitions were refined by classifying 
a small subset of gt\_final descriptions.
The analysts used the classification results and LLM explanations 
to reduce overlaps between the definitions.
Table~\ref{tab:definitions} in the Appendix
details the final definitions used.

\begin{figure}
\small
\centering
\begin{query}[Prompt]
You are a cybersecurity expert with extensive knowledge about scams and abuses. You will help me classify abuse reports given by users, based exclusively on the content of the given TEXT. Do not infer or assume facts that are not described in the TEXT.

The following is a LIST OF DEFINITIONS of abuse classes. Read carefully the list and use it to classify the TEXT at the end, and to answer the QUESTION.

\#\#\# LIST OF DEFINITIONS \#\#\#
$<$definition\_list$>$

TEXT: $<$report\_description\_text$>$

QUESTION: Given the LIST OF DEFINITIONS above, classify the TEXT in one of the following classes: $<$class\_list$>$. Answer only with the name of the class that clearly matches one of these definitions and justify your answer by filling the next JSON structure: \{"answer": "", "reasoning": ""\}

\end{query}
\caption{LLM prompt. Macros in angle brackets 
are replaced with abuse type definitions and the description text.}
\label{fig:query}
\end{figure}

Queries to the LLM use the prompt in Figure~\ref{fig:query}
which contains three input macros that are replaced with
the description to be classified ($<$report\_description\_text$>$),
the list of abuse types ($<$class\_list$>$), and
the abuse type definitions ($<$definition\_list$>$).
The prompt instructs the LLM to take the role of a cybersecurity expert 
and avoid assuming facts not in the
description to minimize hallucinations~\cite{ji2023survey}.
It requests the LLM to produce a JSON with
the selected abuse type and the reasoning for the selection.

\paragraph{\updated{Query approach.}}
\updated{To address the impact of overlapping abuse types among 
different taxonomy levels,
we analyze whether a classification approach that follows the taxonomy's 
hierarchy, either in a top-to-bottom or bottom-to-top manner, can improve 
accuracy by removing overlapping definitions. 
In exchange, a higher number of queries may be needed, 
which may increase the cost and classification delay.
We compare these approaches with the one-query approach used by 
the llm-baseline,
which forces the LLM to reason about overlapping definitions
but minimizes the number of queries.} 
Regardless of the approach used, 
queries use the prompt in Figure~\ref{fig:query}.
Differences between the approaches 
are in the number of queries 
to classify a description and in the definition list
provided in each query. 
In detail, we analyze the following three query approaches:

\emph{(a) One-query.}
This approach uses a single query to the LLM per description. 
Each query provides all the abuse type definitions and 
asks the LLM to select the most appropriate abuse type for the description.
This approach minimizes the number of queries, 
making it the fastest approach and also fairly economical.
But, it forces the LLM to distinguish 
abuse subtypes (e.g., \ttag{investment}) 
from their parents (e.g., \ttag{scam}),
which may be challenging as they semantically overlap.

\emph{(b) Top-to-bottom.}
This approach classifies the input description at each 
taxonomy level from top to bottom.
It performs between one and three queries per description, 
making the classification more specific with each query.
An L1 query is performed for all descriptions by providing only the 
\ttag{abuse} and \ttag{notabuse} definitions.
If the L1 label is \ttag{notabuse}, 
the classification outputs \ttag{notabuse}.
Otherwise, it queries the LLM again with the definitions of the four
L2 abuse types,
plus an additional \ttag{ttb\_other} class
that states that the description does not match the other definitions. 
If the L2 label is \ttag{ttb\_other}, 
the classification outputs \ttag{abuse},
i.e., since none of the more specific children types matches, 
it returns the parent type.
Otherwise, it performs a final query using only the child definitions for 
the assigned L2 label plus the additional \ttag{ttb\_other} class. 
For example, if the L2 label was \ttag{extortion}, only the 5 
\ttag{extortion} children and \ttag{ttb\_other} are included in the query.
The returned L3 label is output as the description's abuse type.
The top-to-bottom approach eases the distinction between a parent and 
its children, but it is more expensive in terms of LLM queries.
Most importantly, errors at the top levels accumulate,
i.e., outputting an accurate label may require correctly classifying the 
description up to three times.

\emph{(c) Bottom-to-top.}
This approach classifies the input description at each 
taxonomy level, from bottom to top.
An L3 query is performed for all descriptions.
It provides the 13 leaf definitions plus an additional \ttag{btt\_other} class
that states that the description does not match the given definitions 
and may not even describe any abuse.
If the L3 label is not \ttag{btt\_other}, 
the classification outputs the L3 label.
Otherwise, an L2 query is performed using the four L2 definitions and 
the additional \ttag{btt\_other}.
If the L2 label is not \ttag{btt\_other},
the classification outputs the L2 label.
Otherwise, a final L1 query is performed using only the 
\ttag{abuse} and \ttag{notabuse} definitions, and 
the classification outputs the assigned L1 label.
This approach is typically the most economical. 
While it performs more queries than the one-query approach, 
the queries are smaller as only a subset of 
definitions is included in each query.
However, it requires distinguishing more classes compared to the 
top-to-bottom approach, which may impact the accuracy.

\paragraph{Models.}
We evaluate six LLM models. 
We download and run locally two open models 
from HuggingFace~\cite{llama3-huggingface}:
Meta-Llama-3-70B-Instruct (llama3 for short) and 
Meta-Llama-3.1-70B-Instruct (llama3.1).
We also evaluate four commercial models from OpenAI accessible via API: 
the older and cheaper gpt-3.5-turbo-0125 (gpt-3.5),
the more recent and more expensive gpt-4-0125-preview (gpt-4), and
two faster models with intermediate cost gpt-4o-2024-05-13 (gpt-4o) and
gpt-4o-mini-2024-07-18 (gpt4-4o-mini).

\paragraph{\updated{Data pre-processing.}}
Prior to classifying a description, we convert its encoding to UTF-8. 
This pre-processing reduces the number of tokens compared to 
encodings such as UTF-16,
which helps keep the description within the maximum model token length 
(e.g., 16K for gpt-3.5) and reduces the query cost and processing time.

\subsection{\updated{Supervised Classifier}}
\label{sec:supervised}

We also build a supervised classifier based on transfer learning.
We select the pre-trained DistilBERT model~\cite{sanh2019distilbert}
available from Hugging Face~\cite{huggingface} and
fine-tune it for 5 epochs with the following default hyper-parameters:
learning\_rate=2e-5 and
weight\_decay=0.01.
We build training and evaluation datasets by
performing a 70/30 split of the descriptions for each abuse type
in gt\_final.
We ensure the datasets are balanced by downsizing the
dominant \ttag{sextortion} and \ttag{extortion} classes
by selecting 1/20 and 1/2 descriptions respectively.
We remove the 5 under-represented abuse types that have
less than 10 descriptions:
\ttag{ransom}, \ttag{unauthwithdrawal}, \ttag{eshop},
\ttag{abuse}, and \ttag{fundsrecovery}.
Thus, the supervised classifier has 14 classes.
The balanced training and evaluation datasets have 2,446 and 1,049
descriptions, respectively.

\begin{table*}[t]
\centering
\scriptsize
\begin{tabular}{lcl|rrr|rrr|rrr}
\hline
\multicolumn{3}{c|}{} & \multicolumn{3}{c|}{\textbf{Weighted Average}} & \multicolumn{3}{c|}{\textbf{Macro Average}} & \multicolumn{3}{c}{\textbf{Cost}} \\
\textbf{Classifier} & \textbf{Definitions} & \textbf{Model} & \textbf{Prec.} & \textbf{Rec.} & \textbf{F1} & \textbf{Prec.} & \textbf{Rec.} & \textbf{F1} & \textbf{Queries} & \textbf{Tokens} & \textbf{USD} \\
\hline
supervised & \N & distillbert-14-bal & 0.99 & 0.99 & 0.99 & 0.72 & 0.75 & 0.73 & - & - & - \\
\updated{llm-baseline} & \N & gpt-3.5 & 0.47 & 0.47 & 0.42 & 0.27 & 0.27 & 0.20 & \textbf{1,049} & \textbf{416,295} & \textbf{\$0.26} \\ one-query & \Y & gpt-3.5 & 0.57 & 0.54 & 0.52 & 0.46 & 0.42 & 0.33 & \textbf{1,049} & 1,927,752 & \$1.02 \\
bottom-to-top & \Y & gpt-3.5 & 0.65 & 0.67 & 0.60 & 0.59 & 0.65 & 0.56 & 1,126 & 1,461,260 & \$0.79 \\ 
top-to-bottom & \Y & gpt-3.5 & 0.85 & 0.82 & 0.81 & 0.63 & 0.72 & 0.65 & 2,871 & 1,993,112 & \$1.13 \\
top-to-bottom & \Y & llama3 & 0.74 & 0.70 & 0.68 & 0.71 & 0.77 & 0.68 & 1,719 & - & - \\
top-to-bottom & \Y & llama3.1 & 0.83 & 0.78 & 0.77 & 0.68 & 0.65 & 0.59 & 2,698 & - & - \\
top-to-bottom & \Y & gpt-4o-mini & 0.84 & 0.83 & 0.82 & 0.79 & 0.80 & 0.76 & 2,837 & 2,074,957 & \$0.40 \\
top-to-bottom & \Y & gpt-4o & 0.95 & 0.92 & 0.92 & 0.79 & 0.73 & 0.71 & 2,830 & 2,090,951 & \$12.76 \\
top-to-bottom & \Y & gpt-4 & \textbf{0.96} & \textbf{0.95} & \textbf{0.95} & \textbf{0.83} & \textbf{0.82} & \textbf{0.80} & 2,901 & 2,218,064 & \$28.08 \\
\hline
\end{tabular}
\caption{Classification results on 1,049 \ba descriptions. 
The top two rows are the supervised and naive LLM baselines.
Our best design in the bottom row more than doubles the F1 score compared 
to the naive LLM baseline in the second row.}
\label{tab:eval_llm}
\end{table*}

\subsection{\updated{Classifier Evaluation on \ba}}
\label{sec:eval}

Table~\ref{tab:eval_llm} summarizes the classification results for each
classifier design on the balanced testing dataset with 1,049 descriptions.
For this, we use both weighted average
(classes are weighted by the number of samples in the class) and
macro average
(all classes have the same weight)
scores.
We use a strict view of accuracy which considers a true positive only if the 
exact same label in the ground truth is output by the classifier.
For example, if the ground truth says \ttag{sextortion} and the classifer 
instead outputs the parent \ttag{extortion} we consider that an error.

The top two rows capture the supervised and naive LLM baselines.
We first compare the naive LLM baseline with the
three query approaches, all using the gpt-3.5 model.
The results show that the naive LLM baseline works worst,
achieving a weighted F1 score of 0.42,
illustrating that a naive LLM-based design does not work and
the importance of the query approach and the abuse definitions
for accurate classification.
Among the three query approaches,
top-to-bottom works best,
achieving a weighted F1 score of 0.81,
followed by bottom-to-top (0.60) and
one-query (0.52).
Since top-to-bottom works best, we compare the six LLM models
using that approach.
The results show that gpt-4 is the best-performing model,
achieving a weighted precision of 0.96, recall of 0.95, and F1 score of 0.95.
The macro average results are lower
due to errors in infrequent abuse types
that disproportionately impact this average.
For example, \ttag{fundsrecovery}, \ttag{abuse}, and \ttag{unauthwithdrawal}
have only 1--2 samples in the testing dataset, so their results are not
very representative but greatly affect the macro average.

The open Llama models are the cheapest, as we run them locally.
But, they achieve the worst weighted average.
The cost of the OpenAI API depends on
the number of queries,
the total size (measured in tokens)
of the input and output text (including the reasoning), and
the model~\cite{openai-pricing}.
Both gpt-4 models are an order of magnitude more costly
than gpt-3.5 and gpt-4o-mini,
and the best-performing gpt-4
is twice as expensive as gpt-4o.
Thus, the most expensive model indeed achieves the best results.

The supervised classifier baseline (distillbert-14-bal)
achieves a nearly perfect
weighted average of 0.99 and 0.73 macro average F1 score.
The results of the supervised classifier are so good that we believe
it may be overfitting the data.
To check this, we next evaluate all classifiers on gt\_bbb.

\begin{table}
\centering
\scriptsize
\resizebox{\columnwidth}{!}{
\begin{tabular}{l|rrr|rrr}
\hline
\multicolumn{1}{c|}{} & \multicolumn{3}{c|}{\textbf{Weighted Average}} & \multicolumn{3}{c}{\textbf{Macro Average}} \\
\textbf{Classifier} & \textbf{Prec.} & \textbf{Rec.} & \textbf{F1} & \textbf{Prec.} & \textbf{Rec.} & \textbf{F1} \\
\hline
distillbert-14-bal & 0.73 & 0.57 & 0.51 & 0.41 & 0.33 & 0.25 \\
\hline
top-to-bottom (llama3) & 0.80 & 0.69 & 0.70 & 0.34 & 0.43 & 0.34 \\
top-to-bottom (llama3.1) & 0.86 & 0.75 & 0.76 & 0.44 & 0.55 & 0.44 \\
top-to-bottom (gpt-3.5) & 0.79 & 0.62 & 0.61 & 0.28 & 0.39 & 0.26 \\
top-to-bottom (gpt-4o-mini) & 0.71 & 0.69 & 0.69 & 0.27 & 0.39 & 0.30 \\ top-to-bottom (gpt-4o) & 0.88 & 0.80 & 0.80 & 0.45 & 0.47 & 0.42 \\ top-to-bottom (gpt-4) & \textbf{0.92} & \textbf{0.87} & \textbf{0.89} & \textbf{0.67} & \textbf{0.76} & \textbf{0.70} \\ \hline
\end{tabular}
}
\caption{Classification results on gt\_bbb.
}
\label{tab:eval_bbb}
\end{table}

\subsection{\updated{Classifier Evaluation on Scam Tracker}}
\label{sec:eval_bbb}

One challenge with classifiers is how well they generalize to 
datasets different from those used to build them.
To examine this issue, we perform an out-of-distribution experiment 
where we apply all classifiers on gt\_bbb. 
One of the 200 descriptions exceeds the 
maximum LLM model token length, so we removed it, leaving 199.
As shown in Table~\ref{tab:eval_bbb}, the best classifier is the LLM-based with top-to-bottom and gpt-4, 
which achieves a weighted average precision of 
0.92, a recall of 0.87, and an F1 score of 0.89.
The F1 score is only 6 percentile points below that achieved on the \ba testing 
dataset, indicating that the model generalizes well. 
In contrast, the supervised classifier performed worst 
with an F1 score of 0.51, confirming that it was overfitting the data
in the \ba dataset.

Most errors for the best classifier happen due to \ttag{investment} reports 
labeled as \ttag{advancefee}.
They are due to the \ttag{investment} reports mentioning upfront fees, 
e.g., ``[...]Then when trade happened he stated that there will be a fee 
that I have to pay for taxes upfront[...]''.
Indeed, most investment scams request additional fees when the user wants 
to withdraw its profits. 
The difference with \ttag{advancefee} is subtle in that in an \ttag{investment}
scam, the victim first makes an investment deposit, and the fee is only requested 
when the victim attempts to withdraw the investment profits.
Instead, in \ttag{advancefee} scams, the fee is requested prior to the 
victim making any deposit.
The \ttag{advancefee} definition used by the LLM classifier
(shown in Table~\ref{tab:definitions}) 
does not explicitly capture this timing constraint.

\subsection{Updating the Taxonomy and Classifier}
\label{sec:updates}

Scam Tracker bundles all cryptocurrency abuse reports into one category,
preventing comparing finer-grained cryptocurrency abuse types.
To address this issue, we apply the
top-to-bottom gpt-4 classifier to the entire Scam Tracker dataset.
We exclude two descriptions that are too large
to be classified by gpt-4.
Column All in Table~\ref{tab:bbb} captures the number of reports of each
type identified.
85\% of the reports are for scams with
half being for \ttag{investment} scams.
Despite its name, Scam Tracker also contains small percentages
of extortions (5.4\%) and unauthorized withdrawal (2\%) reports.
While there is only one \ttag{fundsrecovery} report in gt\_final and
none in gt\_bbb,
the classifier found 16 across the Scam Tracker dataset.
We manually validated that these reports are correctly classified.
This shows that given a good abuse definition the
LLM-based classifier can correctly classify infrequent abuse types.

As described in Section~\ref{sec:taxonomy},
we have designed our hierarchical taxonomy to capture previously unknown
abuse types in the L2 types and the L1 \ttag{abuse} type.
Thus, we manually examine the reports assigned those types
(i.e., 452 \ttag{scam}, 44 \ttag{abuse},
42 \ttag{unauthwithdrawal}, and 16 \ttag{extortion})
to check for new abuse types,
Among the 452 \ttag{scam} reports,
we identify a new abuse type for
\ttag{employment} scams.
We produced a definition for this abuse type and
ran the top-to-bottom gpt-4 classification
on the Scam Tracker descriptions again.
The new classification results identify 56 \ttag{employment} descriptions.
Manual verification identifies 54 as true positives and
two as false positives.
We also examine all descriptions
containing related keywords (e.g., \emph{job}, \emph{employment}),
finding no false negatives.
This process illustrates the ease of extending our taxonomy and LLM-based
classifier with new abuse types.

\subsection{Spam Detection}
\label{sec:spamdetection}

Next, we check how accurately our classifier identifies spam reports. 
For this, we leverage the 287 spam descriptions identified in 
Section~\ref{sec:spam} across the entire \ba dataset.
We classify those descriptions using the L1 query of the 
top-to-bottom gpt-4 classifier.
The classifier achieves a precision of 1.0, a recall of 0.99, and 
an F1 score of 0.99. It correctly identifies 284 (99\%) descriptions as \ttag{notabuse}.
The other three descriptions were considered \ttag{spam\_fundsrecovery} 
in Section~\ref{sec:spam} because their text contains contact IOCs from 
funds recovery services. 
However, the classifier correctly determines that they are valid abuse
reports where victims describe how they were scammed by a funds recovery 
service (i.e., \ttag{fundsrecovery}). 

We also examine the 140 Scam Tracker reports classified as \ttag{notabuse}.
Among those, we identify 9 \ttag{spam\_fundsrecovery} reports
(0.4\% of all Scam Tracker reports)
that Scam Tracker's manual validation failed to filter
but are correctly identified as invalid reports by our classifier.

The results show that our classifier can accurately filter 
spam reports, 
including some missed by manual validation. 
Furthermore, our LLM-based classifier 
correctly distinguishes useless \ttag{spam\_fundsrecovery} advertisements from 
valid \ttag{fundsrecovery} scam reports by victims.

\section{Abuse Type Revenue Estimation}
\label{sec:estimation}

This section answers RQ3 by analyzing the financial impact of 
different abuse types.
Two approaches exist for estimating the financial impact of
cryptocurrency abuse.
The first one adds losses disclosed by victims in abuse reports.
Unfortunately, victims often do not report the abuse 
suffered~\cite{smith2007consumer,citizens_advice_toolkit,tmj4_scam_reporting},
making this estimation a conservative lower bound. 
The alternative is analyzing transactions in the blockchain ledger
that deposit funds to reported malicious addresses.
This approach enables observing deposits 
from victims who did not report an attack.
However, it may overestimate if benign addresses (e.g., from exchanges) 
are mistakenly included
and if addresses are used in multiple abuse types.

\begin{table}[t]
\centering
\resizebox{\columnwidth}{!}{
\begin{tabular}{l|rr|rrr}
\hline
\multicolumn{1}{c|}{} & \multicolumn{2}{c|}{\textbf{Reports}} & \multicolumn{3}{c}{\textbf{Loss}} \\
\textbf{Abuse} & \textbf{All} & \textbf{w/Loss} & \textbf{Total} & \textbf{Avg.} & \textbf{Med.} \\
\hline

investment & 1,161 (50\%) & 957 (61\%) & \$22,502,025 & \$19,382 & \$1,300 \\
scam & 452 (19\%) & 243 (16\%) & \$3,072,803 & \$6,798 & \$100 \\
advancefee & 175 (8\%) & 127 (8\%) & \$1,191,216 & \$6,807 & \$226 \\
notabuse & 140 (6\%) & 69 (4\%) & \$1,105,560 & \$7,897 & \$0 \\
sextortion & 104 (4\%) & 0 (0\%) & \$0 & \$0 & \$0 \\
giveaway & 75 (3\%) & 38 (2\%) & \$149,831 & \$1,998 & \$20 \\
romance & 57 (2\%) & 34 (2\%) & \$2,181,882 & \$38,279 & \$1,500 \\
abuse & 44 (2\%) & 21 (1\%) & \$211,436 & \$4,805 & \$0 \\
unauthwith. & 42 (2\%) & 27 (2\%) & \$224,270 & \$5,340 & \$220 \\
eshop & 32 (1\%) & 28 (2\%) & \$99,226 & \$3,101 & \$600 \\
fundsrecovery & 16 (1\%) & 12 (1\%) & \$91,077 & \$5,692 & \$1,710 \\
extortion & 16 (1\%) & 2 (0\%) & \$2,552 & \$160 & \$0 \\
reputation & 3 (0\%) & 0 (0\%) & \$0 & \$0 & \$0 \\
harm & 2 (0\%) & 2 (0\%) & \$1,100 & \$550 & \$1,000 \\
\hline
All & 2,319 (100\%) & 1,560 (67\%) & \$30,832,979 & \$13,297 & \$500 \\

\hline
\end{tabular}
}
\caption{Scam Tracker reports classified into each abuse type,
the subset reporting a financial loss,
the total loss in reports of the type, and
the average and median loss per report.
}
\label{tab:bbb}
\end{table}

\subsection{Victim-Reported Financial Losses}
\label{sec:losses}

Scam Tracker reports contain an optional field with victim-reported losses 
in US dollars.
While Scam Tracker bundles all cryptocurrency abuse reports in 
one category, 
we can use the classification results on the 
Scam Tracker descriptions, introduced in Section~\ref{sec:updates}, 
to compare the financial impact of different cryptocurrency abuse types.

Table~\ref{tab:bbb} summarizes the results. 
Overall, 67\% of the cryptocurrency abuse reports include a financial loss, 
the total reported loss was \$30.8M, 
with an average loss of \$13K and a median of \$500 per report.
However, there are significant differences across abuse types.
Financial loss is reported in 61\% of \ttag{investment} reports, 
nearly four times larger than any other abuse type.
Thus, \ttag{investment} scam targets are most likely to become victims.
The highest median loss is for 
\ttag{fundsrecovery} scams (\$1,710), 
followed by \ttag{romance} scams (\$1,500), and 
\ttag{investment} scams (\$1,300).
The largest average is for \ttag{romance} (\$38,279) and 
\ttag{investment} (\$19,382) as these categories contain the 
reports with the largest losses.
In contrast, extortions seem particularly inefficient at obtaining 
revenue from victims.
For example, no losses are reported for \ttag{sextortion} and \ttag{reputation}.
Extortions typically leverage massive email campaigns to reach large numbers 
of targets to compensate for their low conversion rate.
Another category with small median losses is 
\ttag{giveaway} (\$20) where victims choose how much to send, and may 
conservatively select small test amounts. 
Of the 9 \ttag{spam\_fundsrecovery} advertisements 
7 report a loss, for a total of \$548,457.
Thus, spam reports can pollute financial estimations by reporting 
fictitious losses.

\subsection{Estimation from Address Deposits}
\label{sec:tagging}

We propose a three-step approach
to estimate revenues per abuse type
from the transactions in the blockchain.
\updated{First, we use our LLM-based classifier to assign an abuse type 
to each report.
Second, 
we assign each reported address an abuse type
by selecting the most common label
from its now-labeled abuse reports.
Third, we apply \wyb~\cite{wyb} to 
estimate the amount of BTC received by each address
from the first block of the Bitcoin blockchain,
up to block 838,800 (mined on April 11, 2024).
As recommended by Gomez et al.~\cite{btcestimations}, 
we filter out self-deposits to avoid double-counting
and use the BTC conversion rate to USD on the day of each deposit.
Finally, we aggregate the revenue for addresses assigned the same abuse type.}

Our tagging procedure assigns an abuse type to an address
by performing a majority vote on the classifier results for its reports.
Given an address, it iterates on all the reports for the address.
For each report, it obtains the abuse type output by the classifier
for the description text.
If the report's abuse type is \ttag{notabuse}, the report is ignored.
Otherwise, it increases a counter for the abuse type.
After processing all reports for the address, the abuse type with the
highest counter is output as the abuse type for the address.
If all reports were \ttag{notabuse}, no abuse type is output.

\paragraph{Benign address impact.}
\updated{Removing the benign addresses is fundamental for
a correct estimation because those addresses may receive large 
amounts of bitcoins unrelated to cybercrime.
Failure to exclude them can hugely overestimate cybercrime revenue.
To evaluate this impact, we first compute the total revenue 
across all addresses, including the benign ones.
The \numaddronchain addresses with transactions have collectively received
\estBtc BTCs, corresponding to \estUsd.
Then, we repeat the estimation excluding the 91 benign addresses,
which reduces the estimation to \estNoInternalBtc BTC (\estNoInternalUsd).
Thus, even if benign addresses represent only 0.1\% of reported addresses,
their impact on cybercrime revenue estimation is overwhelming.
Excluding them reduces the estimation by 60\%
from \$2.4 trillion to \$950 billion.
The latter is still likely an overestimation,
since additional internal service addresses,
unknown to the four tagging sources used,
could remain unfiltered.}
 
\begin{table}
\centering
\resizebox{\columnwidth}{!}{
\begin{tabular}{lr|rr|rr}
\hline
\multicolumn{2}{c|}{} & \multicolumn{2}{c|}{\textbf{Tagged Addresses}} & \multicolumn{2}{c}{\textbf{Revenue}} \\
\textbf{Abuse} & \textbf{Rep.} & \textbf{All} & \textbf{w/Dep.} & \textbf{BTC} & \textbf{USD} \\
\hline

investment & 209 & 202 & 201 & 8,060.2558 & \$199,717,207 \\
scam & 323 & 236 & 229 & 4,392.8819 & \$121,183,928 \\
unauthwithdrawal & 60 & 57 & 57 & 2,729.4449 & \$65,097,775 \\
abuse & 106 & 49 & 48 & 1,651.4201 & \$20,317,573 \\
giveaway & 103 & 102 & 102 & 1,135.9124 & \$11,151,614 \\
romance & 17 & 17 & 17 & 359.4312 & \$10,951,634 \\
advancefee & 46 & 42 & 42 & 390.6354 & \$10,449,521 \\
extortion & 1,979 & 915 & 462 & 362.5466 & \$7,099,341 \\
eshop & 26 & 25 & 25 & 80.4185 & \$2,610,870 \\
sextortion & 2,141 & 767 & 698 & 285.4095 & \$2,474,572 \\
fundsrecovery & 5 & 5 & 5 & 50.8731 & \$1,542,469 \\
ransom & 111 & 62 & 35 & 19.7539 & \$453,233 \\
clipper & 8 & 7 & 7 & 10.0543 & \$350,226 \\
harm & 2 & 1 & 1 & 4.5950 & \$271,174 \\
deleteransom & 49 & 24 & 24 & 18.9006 & \$174,852 \\
ddos & 1 & 1 & 1 & 0.0865 & \$4,558 \\
reputation & 27 & 4 & 4 & 0.1691 & \$1,533 \\
bomb & 2 & 2 & 2 & 0.0000 & \$0 \\
\hline
All abuse & 5,215 & 2,518 & 1,960 & 19,552.7888 & \$453,852,080 \\

\hline
\end{tabular}
}
\caption{Revenue estimation for 1,960 addresses tagged 
using the classification of 5K \ba descriptions.
}
\label{tab:estimation}
\end{table}

\paragraph{\BA estimation.}
Classifying all 224K \ba descriptions would
cost roughly \$5,000, a budget we currently do not have.
Instead, we classify 5K descriptions using the 
LLM-based top-to-bottom with gpt-4 classifier, with a cost of \$109.
Those 5K descriptions are used in 7,322 reports for 3,668 addresses.
We exclude 33 benign addresses and apply our tagging procedure to the rest, 
generating tags for 2,518 (69\%) addresses.
The rest of the addresses are not tagged
because their reports are classified as \ttag{notabuse}.
Table~\ref{tab:estimation} presents the estimation 
for the 1,960 tagged addresses with at least one deposit. 
Due to the fluctuating BTC conversion rate, 
abuse types with smaller BTC revenue
can have larger USD revenue
(e.g., \ttag{eshop} vs \ttag{sextortion}).
Similar to the estimation from victim-reported lossses,
\ttag{investment} brings the most profit, and extortions 
like \ttag{bomb}, \ttag{reputation}, or \ttag{ddos}
are fairly ineffective.
However, we identify \$2.4M \ttag{sextortion} revenue
due to this abuse type having the 
largest number of tagged addresses (698).
This again points to extortions having a low conversion rate 
but obtaining revenue through massive spam campaigns.

\paragraph{Scam Tracker.}
Although Scam Tracker does not have a field to report cryptocurrency addresses, 
there are 71 reports mentioning 96 Bitcoin addresses in the description text.
Using these 71 reports, we can compare the victim-reported losses with the 
estimation from the deposits to the mentioned addresses.
While victims disclosed losses of \$520K in the 71 reports, 
the revenue computed from the deposits to the 96 addresses 
is \$15.1M, 29 times higher.
The extreme case is one \ttag{unauthwithdrawal} address 
with \$76K victim-reported losses that has received \$6.6M dollars,
i.e., victim-reported losses capture only 1.1\% of the total revenue.
Table~\ref{tab:bbb-estimation} in the Appendix details the comparison
of both estimation approaches.
Our results show that the victim-reported losses provide 
a very conservative lower bound of the revenue achieved by the 
reported cybercriminal operations, 
with the real revenue being up to 29 times higher.
\section{Discussion}
\label{sec:discussion}

\updated{This section discusses the utility of our results, 
including how they generalize to other services and blockchains, 
how the different classifier types compare, and
the limitations of our approach.}

\subsection{\updated{Utility of the Results}}
\label{sec:utility}

Our results are useful in different ways.
\updated{First, our classifier can be used to assign abuse types to 
cryptocurrency addresses.
Those labeled addresses are crucial to blockchain analysis platforms, 
enabling services to identify cybercriminal money-laundering transactions.}

Second, our classifier, 
can be used by abuse reporting services 
to filter invalid reports.
These include spam reports, which we have shown that, if left unchecked, 
will eventually flood an abuse reporting service.
Furthermore, filtering spam protects victims from advertisements 
of funds recovery scams, which we have measured to
have the highest median loss at \$1,700 per report. 
Beyond spam, our classifier also identifies other useless reports that do not report abuse and should also be filtered.
We observe that Scam Tracker's manual validation
prevents spam reports from flooding their database,
but 0.4\% of reports are still unfiltered spam,
possibly due to analyst fatigue or inconsistency.
Such inconsistencies and the increasing number of scams 
further motivate the use of classifiers to automate the filtering. 

Third, our classifier also enables valid abuse reports to be classified
into fine-grained abuse types.
This has enabled us to analyze the financial impact of different abuse types. 
While abuse reporting services already quantify financial impact 
based on victim-reported losses, 
we have shown that victim-reported losses largely underestimate the 
cybercriminal's revenue, which we have measured to be 29 times higher 
using address deposits.
Scam Tracker does not collect blockchain addresses 
so it cannot estimate address deposits. 
ChainAbuse, 
which acquired BitcoinAbuse~\cite{baAcquisition}, 
collects blockchain addresses in the reports, 
but leverages user-selected abuse types,
which we have shown to be inaccurate.
Our results show that the report classification should be automated 
because victims do not care about it
(i.e., often selecting the first option) and 
make many mistakes likely due to lack of experience.
One could be tempted to think that simply giving the abuse reports to the 
LLM is enough to classify them. 
However, the naive LLM usage baseline, which does precisely that,
only achieves a 0.42 F1 score compared to 0.95 for our best classifier design.

Fourth, our measurements show that most reports are for extortion
and predominantly \ttag{sextortion}. 
This is due to massive email campaigns used to compensate for their low 
conversion rate. 
One defense is for email servers to scan messages for 
blockchain addresses and check those addresses using blocklists of 
reported addresses.
Our measurements also show victims fall for investment scams four times
more often than for other abuse types and their median loss is the 
second largest at \$1,500 per report.
Thus, novel defenses against investment scams are an absolute priority.

\updated{Fifth}, we provide a public dataset of nearly 20,000 abuse descriptions
annotated with the 19 abuse types in our taxonomy. 
We believe this dataset is unique due to its size, 
the wealth of cryptocurrency abuse types it covers, and 
the quality of the labels. 
\updated{Beyond training and evaluating abuse report classification approaches,
this dataset could also be used to train generative AI models
capable of creating synthetic abuse reports.
Those synthetic adversarial reports could then be used in training analysts
and for evaluating defenses against pollution in abuse reporting services.}

\paragraph{Report value.}
The acquisition of \BA by ChainAbuse indicates that, 
despite the pollution we identify,
the \BA reports were considered valuable.
To further establish their value, we check the reported 
addresses against the known malicious addresses in \wyb~\cite{wyb},
which come from published academic datasets or are mentioned in 
the blogs of security vendors. 
We find 875 (0.9\%) known malicious addresses, 
an order of magnitude more than the 91 (0.1\%) known benign addresses 
identified. 
This indicates that the majority of reported addresses 
are malicious and could be used to warn other users who 
encounter them.

\subsection{\updated{Generalization of the Results}}
\label{sec:generalization}

\updated{This section discusses how our results generalize to other 
blockchains and other abuse reporting datasets.

\paragraph{Blockchain-agnostic design.}
We have designed our approach to be agnostic of the blockchain used in the 
attacks, so it can analyze reports of abuse on any blockchain. 
Our LLM classifier identifies the abuse type based on the textual 
description the victim provides without any references in the prompt to 
a specific blockchain. 
While \ba focused on abuse reports for Bitcoin addresses, 
our ScamTracker dataset contains cryptocurrency abuse reports
on a variety of blockchains beyond Bitcoin,
such as Ethereum, Tron, Ripple, Dogecoin, and Dashcoin.
The only blockchain-specific part of the paper is the revenue estimation 
from address deposits in Section~\ref{sec:tagging}.
To estimate revenue from address deposits beyond Bitcoin,
we could leverage a more general blockchain analysis 
platform such as GraphSense~\cite{graphsense} or the APIs of blockchain 
explorer services to collect transaction data
from other blockchains~\cite{blockchain-com,blockchair,etherscan}.

\paragraph{Service-agnostic design.}
We have designed our approach to be agnostic of the reporting service 
providing the abuse reports. 
Our LLM classifier identifies the abuse type based exclusively on the textual 
description the victim provides without using any 
service-specific report metadata.
We have demonstrated our approach on reports obtained from \ba and ScamTracker, 
but we could equally apply it to reports from other abuse reporting services 
that collect victims' textual 
descriptions~\cite{chainabuse,scamwatch,reportfraud}.
Furthermore, our classifier could be applied to less noisy datasets, 
e.g., victim complaints directly filed with LEAs.
Since LEAs may not be able to submit complaints to commercial LLMs,
we could deploy our classifier on-site using the open models, 
albeit at reduced accuracy. 

The pollution we identified may also 
affect other abuse reporting services. 
For example, when ChainAbuse acquired \ba, 
it added the \ba reports to its database.
To check whether they removed 
some of the pollution we observed, 
we queried the 91 benign addresses 
through the ChainAbuse webpage in October 2025. 
ChainAbuse had removed all reports for 61 addresses, 
but another 30 still appear in their database.
The reports for those addresses still include some 
funds recovery and terrorism spam reports. 
Among others, ChainAbuse contains reports for 
an address\footnote{1F1tAaz5x1HUXrCNLbtMDqcw6o5GNn4xqX}
known to be used by the FBI to seize SilkRoad's funds~\cite{fbi}
and addresses that hold reserves of exchanges like
Bitfinex\footnote{1KYiKJEfdJtap9QX2v9BXJMpz2SfU4pgZw} and 
BitMex\footnote{3BMEX4AJFi3JbU4vdBXMaNTfUH28H6WDWq},
publicly stated by themselves~\cite{bitfinex,bitmex}.
This shows that identifying benign addresses and spam reports is challenging 
even for specialized commercial services.

\paragraph{Beyond cryptocurrency abuse.}
Most abuse types in our taxonomy are not specific to cryptocurrencies.
For example, \ttag{investment} scams could request investments through regular
payment mechanisms such as bank transfers or credit card payments.
Thus, we believe our classifier could be generalized to also handle 
non-cryptocurrency abuse reports.
The challenge would be generalizing the definitions 
used by the classifier to be cryptocurrency-independent.

\paragraph{Beyond abuse reports.}
Our LLM-based classifier could also be applied in other 
threat intelligence workflows. 
For example, it could be extended to examine threat reports
published by cybersecurity vendors, 
in order to label cryptocurrency addresses appearing in the reports. 
}

\subsection{\updated{Classifier Comparison}}
\label{sec:comparison}

\updated{We have compared our unsupervised LLM-based classifier with a 
supervised classifier baseline.
In this section, we discuss the advantages and disadvantages of 
both types of classifiers.

The main advantages of the supervised classifier are 
its high accuracy on similar data, high scalability, 
lower environmental impact, and lower cost.
As shown in Table~\ref{tab:eval_llm}, 
the supervised classifier achieves the highest 0.95 F1 score on 1,049 \ba test 
descriptions from gt\_final,
compared to 0.92 for the best LLM-based classifier. 
This likely happens because gt\_final contains many syntactically similar 
reports from the same cluster, which the supervised classifier accurately models.
However, the supervised classifier performed worst
when applied to abuse reports from a different reporting service
(during the out-of-distribution experiment),
achieving a 0.51 F1 score
compared to 0.61--0.89 for the LLM-based classifiers. 
Thus, the supervised classifier does not generalize well to abuse reports 
different from those in the training data. 
Scalability is also an advantage; 
once trained, the supervised classifier labels the 1,049 reports in
minutes compared to a few hours for the best LLM-based classifier,
which needs to frequently query the gpt-4 cloud API.
Cost, both in financial and energy consumption terms, 
is also a clear advantage.
The supervised classifier doesn't generate commercial API costs,
and the environmental impact due to energy consumption is arguably 
much smaller than using an LLM (although hard to quantify).

On the other hand, the unsupervised LLM-based classifier has 
advantages in terms of generalization on new data,
handling of underrepresented classes, 
ease of updating, and 
explainability.
The generalization advantage has already been described above. 
Since it does not require training, the LLM-based classifier 
can easily identify underrepresented classes. 
For example, gt\_final contains only one \ttag{fundsrecovery} 
description and gt\_bbb contains none, 
making it extremely hard for the supervised classifier to model that abuse type.
But the LLM-based classifier 
is able to leverage the \ttag{fundsrecovery} definition 
to correctly identify 16 descriptions of that abuse type 
across the entire Scam Tracker dataset.
The LLM-based classifier is also easier to update; 
adding a new abuse type requires only a new definition,
as demonstrated with \ttag{employment} scams in Section~\ref{sec:updates}.
In contrast, the supervised classifier would require collecting multiple 
reports of the new abuse type and re-training the classifier, 
which took 10 hours on our server.
Finally, the LLM-based classifier can generate a justification for its decision, 
which we found extremely useful for refining the abuse type definitions.
In contrast, the supervised classifier is largely a black-box, 
so its errors are harder to analyze.

We believe that the unsupervised LLM-based classifier is currently the best option
because generalization to previously unseen data is critical:
it's difficult to predict what kind of 
abuse will be reported, and attackers frequently update their tactics.
Furthermore, open-source LLMs 
(e.g., llama3.1 or the recently released gpt-oss~\cite{gpt-oss}) 
can be installed locally,
eliminating the cost advantage of the supervised classifier
and reducing the scalability gap.
Using open models does not necessarily affect accuracy, as 
the llama open models performed better than the 
supervised classifier on the ScamTracker reports.
}

\subsection{Limitations}
\label{sec:limitations}

\paragraph{Evasion.}
\updated{We have proposed a novel LLM-based approach to defend against 
pollution in crowd-sourced abuse reporting services. 
However, attackers could abuse LLMs to bypass such defenses. 
For instance, while current attackers typically generate large volumes of 
spam abuse reports using a few templates, 
LLMs could enable them to create unique, convincing reports that are harder 
to detect -- both by clustering methods and human analysts.

To counter this, we could leverage generative AI to produce synthetic abuse 
reports for training analysts, helping them recognize subtle spam patterns. 
Additionally, future work could explore using generative AI to create 
fake abuse reports that specifically target and challenge LLM-based defenses, 
improving their ability to detect and filter out polluted reports.}

\paragraph{Revenue estimations.}
\updated{Our methodology for estimating revenue 
from Bitcoin address deposits follows state-of-the-art 
recommendations~\cite{btcestimations}.
Still, some risks remain. 
First, we only observe a limited subset of payment addresses of any abuse type,
which can introduce significant underestimation 
(up to 30 times in a case study~\cite{btcestimations}).
Overestimation can also be introduced due to 
unidentified benign addresses and 
the reuse of malicious addresses for different abuse types.
The subset of 3,635 addresses used for the estimation in 
Section~\ref{sec:tagging} does not contain any outliers in terms of 
large amounts of deposits and thus is likely free of benign addresses.
In contrast, when we examine the entire \numaddr addresses in \ba, there are 
some addresses with hundreds of millions of dollars in deposits that are not 
identified as benign by our four 
sources~\cite{glasschain,arkham,graphsense,wyb}, but may likely be. 
Another possible source of overestimation is that we assume that all deposits 
to a reported address come from victims and correspond to the same abuse type.
Instead, an attacker could reuse an address for different purposes.
However, this is less likely as fresh Bitcoin addresses 
can be generated for free,
and reusing addresses facilitates tracking~\cite{fistfulMeiklejohn}.
It is possible to apply value-based and time-based filters to limit 
the deposits included in the estimation.
However, such filters suffer from limited visibility, 
i.e., the amount and time ranges are estimated 
based on the limited visibility of an operation's addresses.
Furthermore, separate filters are needed for 
each malicious operation, which is unfeasible with large datasets such as ours.

Given the limited visibility of our datasets, 
the use of only a subset of the data for estimation, and 
the usage of a conservative estimation methodology, 
the absolute revenue amounts provided in Table~\ref{tab:estimation} 
likely suffer from significant underestimation. 
However, the main value is arguably the relative
comparison of revenue between abuse types,
which identifies abuse types with larger financial impact on victims,
such as \ttag{investment} and \ttag{fundsrecovery} scams.
}

\paragraph{Taxonomy completeness.}
\updated{Any abuse taxonomy is necessarily incomplete.}
Two abuse types analyzed in prior work that we have not observed are
cryptojacking which typically mines Monero~\cite{tekiner2021sok} 
and initial coin offering (ICO) scams that mostly 
affect Ethereum~\cite{tiwari2020future}.
Others appear with different names such as 
Ponzi schemes~\cite{ponziBartoletti} (\ttag{investment})
and tech support scams~\cite{acharya2024conning} (\ttag{fundsrecovery}).
As shown in Section~\ref{sec:updates},
our classifier is easily extensible by adding new abuse definitions.

\paragraph{Data bias.}
\updated{Our work is limited by the available data,
i.e., 290K abuse reports collected from two services.
Pollution may affect other services differently, 
e.g., depending on their validation policies.
We rely on historical data from June 2023 for \ba and May 2024 for ScamTracker.
While our unsupervised LLM-based classifier does not require 
retraining, it may need updates in the long term,
e.g., adding emerging abuse types to the taxonomy
or adjusting abuse type definitions to capture new variations.}

\paragraph{LLM bias and limitations.}
\updated{
Our classifier may suffer from LLM bias. 
For example, it is possible that the LLM training data contains 
information that hampers the classification.
To mitigate this, we have designed the prompt to include our own 
abuse type definitions, avoiding possibly contradictory definitions present 
in the LLM training data.
While we aimed to make our prompt model-agnostic, 
we spent most of our time evaluating gpt4. 
Thus, our prompt may be optimized for this model, 
in which case other models would achieve a lower accuracy than they could.
The LLM may also suffer from socio-linguistic bias,
such as treating informal speech used by some abuse reporters 
as less credible and thus flagging those reports as \ttag{notabuse}.
Our LLM-based classifier currently uses a zero-shot prompt. 
Using a few-shot prompt instead could improve the accuracy, 
although the diversity of abuse reports in our datasets 
makes it hard to select examples that help the LLM generalize better.
To improve classification accuracy, we could also fine-tune an LLM 
using our ground truth. 
In exchange, the classifier would be heavier to update,
as it would require examples of the new abuse type and re-training, 
similar to a supervised classifier.
}

\paragraph{Cost/Scalability.}
\updated{
The cost and scalability of commercial LLMs may limit the classification 
of large datasets, 
e.g., we estimated the cost of classifying 224K abuse reports using 
gpt4 at roughly \$5,000, preventing us from classifying our whole dataset.
However, this is still more affordable than the cost of a human analyst,
which, according to our GT experience, can manually 
classify roughly 200 reports per day.
Furthermore, the cost of commercial models is dropping quickly 
as newer models appear,
e.g., gpt-4o-mini costs an order of magnitude less than gpt-4.
Open-source models are available that can be installed locally,
removing the API costs. 
While open-source models perform worse than commercial ones in our evaluation, 
their performance keeps improving,
e.g., the recently released open-source gpt-oss model achieves
performance similar to gpt4o-mini on core reasoning benchmarks~\cite{gpt-oss}. 
Scalability is also a concern for larger datasets, as LLM queries can take 
from seconds to minutes, depending on available hardware and whether 
the model is installed locally or queried through an API.
Supervised classifiers may be a better option for fast classification of 
large datasets, as discussed in Section~\ref{sec:comparison}. 
}

\section{Related Work}
\label{sec:related}

\paragraph{Cryptocurrency abuse.}
Previous works have analyzed a wealth of cryptocurrency abuse types, including 
ransomware~\cite{bitiodineSpagnoulo,behindLiao,economicConti,trackingHuang,ransomwareClouston},
sextortion~\cite{spamsPaquetClouston},
scams~\cite{ponziBartoletti,fistfulMeiklejohn,xia2020characterizing,bartoletti2021cryptocurrency,li2023giveaway},
clippers~\cite{wyb},
cryptojacking~\cite{huang2014botcoin,tekiner2021sok},
hidden marketplaces~\cite{travelingChristin,cybercriminalLee,dreadRon}, 
human trafficking~\cite{backpagePortnoff},
money laundering~\cite{inquiryMoser}, and 
thefts~\cite{fistfulMeiklejohn}.
These works analyze a single abuse type, 
whereas our work analyzes 19.
Other works compare the revenue obtained by 
different abuse types~\cite{btcestimations,rosenquist2024dark}.
These works require addresses to be tagged with an abuse type,
which we show can be achieved by applying a majority vote on the
results of our classifier. 

\paragraph{Bitcoin analysis.}
Another line of work develops techniques for analyzing Bitcoin flows,
such as multi-input (or co-spending) clustering~\cite{bitcoin,evaluatingAndroulaki,quantitativeRon,fistfulMeiklejohn} 
and change address heuristics~\cite{fistfulMeiklejohn,evaluatingAndroulaki,cookieGoldfeder,automaticErmilov,kappos22peel}. 
Those techniques are implemented by Bitcoin analysis 
platforms~\cite{bitiodineSpagnoulo,blocksci,graphsense,wyb}.
We build our transaction analysis on \wyb~\cite{wyb}. 
 
\paragraph{\ba.}
Previous works analyzing \ba report on the 
presence of heavy reporters~\cite{choi2022large}, 
the concentration of revenue on a few addresses~\cite{buil2022offending}, and
the revenue per abuse type~\cite{rosenquist2024dark}.
However, they did not consider data pollution: 
they did not filter out spam reports,
considered all reported addresses to be malicious
(and thus their deposits to be cybercriminal revenue), and
assumed reporter-selected abuse types were correct.
Our results cast a shadow over their conclusions.

\section{Conclusion}
\label{sec:conclusion}

We have analyzed 289K abuse reports from two services to  
answer three important research questions.
First, we have measured the extent and impact of data pollution.
For this, we have built a public dataset of 19,443 abuse descriptions 
labeled with 19 popular abuse types and have used it to show that, 
if left unchecked,
spam reports will eventually flood an abuse reporting service;
that a small fraction of reported addresses may be benign (0.1\%) but 
those are responsible for a majority (60\%) of the deposited funds; and 
that user-reported abuse types cannot be trusted.
Second, we have shown that the identification of useful reports and 
their classification into abuse types can be automated.
We have presented an LLM-based classifier 
that, on previously unseen data, achieves an F1 score of 0.89
compared to 0.55 for a baseline supervised classifier.
And, it identifies spam reports with a 0.99 F1 score.
Finally, we have compared the financial impact of different cryptocurrency
abuse types showing that victim-reported losses largely under-estimate,
with the revenue estimation
from deposit transactions being 29 times higher.
We show that \ttag{investment} scams have the highest financial impact,
that \ttag{fundsrecovery} and \ttag{romance} scams have the highest 
revenue per report, and 
that extortions have a very low conversion rate.
 
\section*{Funding}
This work was partially funded by
the Spanish Government MCIN/AEI/10.13039/501100011033/
through grants
TED2021-132464B-I00 (PRODIGY),
PID2022-142290OB-I00 (ESPADA), 
PRE2019-088472, and
PREP2022-000165.
The above grants are co-funded by
European Union ESF, EIE, and NextGeneration funds.
This work has also been partially supported by the European commission under SHASAI (Secure Hardware and Software for AI Systems, Grant agreement ID 101225866) project.

\bibliographystyle{IEEEtran}
\bibliography{bibliography/paper}

@inproceedings{wyb,
  author = {Gomez, Gibran and Moreno-Sanchez, Pedro and Caballero, Juan},
  title = {{Watch Your Back: Identifying Cybercrime Financial Relationships in
           Bitcoin through Back-and-Forth Exploration}},
  booktitle = {ACM SIGSAC Conference on Computer and Communications Security},
  year = {2022},
}

@inproceedings{btcestimations,
  author = {Gibran Gomez and Kevin van Liebergen and Juan Caballero},
  title = {{Cybercrime Bitcoin Revenue Estimations: Quantifying the Impact of Methodology and Coverage}},
  booktitle = {ACM SIGSAC Conference on Computer and Communications Security},
  year = {2023},
}

@inproceedings{blocksci,
  author = {Kalodner, Harry and Malte Möser and Kevin Lee and Steven Goldfeder
            and Martin Plattner and Alishah Chator and Arvind Narayanan},
  title = {{BlockSci: Design and Applications of a Blockchain Analysis Platform}
           },
  booktitle = {USENIX Security Symposium},
  year = {2020},
}

@misc{bitcoinabuse,
  title = {{Bitcoin Abuse}},
  note = {\url{https://www.bitcoinabuse.com}},
  year = {2025},
}

@misc{chainabuse,
  title = {ChainAbuse},
  note = {\url{https://chainabuse.com}},
  year = {2025},
}

@misc{scamwatch,
  title = {ScamWatch},
  note = {\url{https://www.scamwatch.gov.au}},
  year = {2025},
}

@misc{scamtracker,
  title = {ScamTracker},
  note = {\url{https://www.bbb.org/scamtracker}},
  year = {2025},
}

@misc{reportfraud,
  title = {ReportFraud},
  note = {\url{https://reportfraud.ftc.gov}},
  year = {2025},
}

@inproceedings{evaluatingAndroulaki,
  author = {Androulaki, Elli and Karame, Ghassan O. and Roeschlin, Marc and Scherer, Tobias and Capkun, Srdjan},
  title = {{Evaluating User Privacy in Bitcoin}},
  booktitle = {Financial Cryptography and Data Security},
  year = {2013},
}

@article{cookieGoldfeder,
  title = {When the cookie meets the blockchain: Privacy risks of web payments via cryptocurrencies},
  author = {Steven Goldfeder and Harry A. Kalodner and Dillon Reisman and Arvind Narayanan},
  journal = {PoPETs},
  year = {2018},
  volume = {2018},
  pages = {179-199}
}

@inproceedings{automaticErmilov,
  title = {{Automatic Bitcoin Address Clustering}},
  author = {Ermilov, Dmitry and Panov, Maxim and Yanovich, Yury},
  booktitle = {IEEE International Conference on Machine Learning and Applications },
  year = {2017},
}

@unpublished{bitcoin,
  author = {Satoshi Nakamoto},
  year = {2008},
  title = {Bitcoin: A Peer-to-Peer Electronic Cash System},
  note = {https://bitcoin.org/bitcoin.pdf},
}

@inproceedings{graphsense,
  title = {{O Bitcoin Where Art Thou? Insight into Large-Scale Transaction Graphs}},
  author = {Bernhard Haslhofer and Roman Karl and Erwin Filtz},
  booktitle = {SEMANTiCS Conference},
  year = {2016}
}

@inproceedings{fistfulMeiklejohn,
  author = {Meiklejohn, Sarah and Pomarole, Marjori and Jordan, Grant and
            Levchenko, Kirill and McCoy, Damon and Voelker, Geoffrey M. and
            Savage, Stefan},
  title = {{A Fistful of Bitcoins: Characterizing Payments among Men with No
           Names}},
  booktitle = {Internet Measurement Conference},
  year = {2013},
}

@inproceedings{travelingChristin,
  title = {{Traveling the Silk Road: A measurement analysis of a large anonymous online marketplace}},
  author = {Nicolas Christin},
  booktitle = {The World Wide Web Conference},
  year={2013}
}

@inproceedings{quantitativeRon,
  author = {Ron, Dorit and Shamir, Adi},
  title = {{Quantitative Analysis of the Full Bitcoin Transaction Graph}},
  booktitle = {Financial Cryptography and Data Security},
  year = {2013},
}

@inproceedings{inquiryMoser,
  author = {M. {Möser} and R. {Böhme} and D. {Breuker}},
  booktitle = {APWG eCrime Researchers Summit},
  title = {{An Inquiry into Money Laundering Tools in the Bitcoin Ecosystem}},
  month = {September},
  year = {2013},
}

@inproceedings{bitiodineSpagnoulo,
  author = {Spagnuolo, Michele and Maggi, Federico and Zanero, Stefano},
  title = {{BitIodine: Extracting Intelligence from the Bitcoin Network}},
  booktitle = {Financial Cryptography and Data Security},
  year = {2014},
}

@inproceedings{dreadRon,
  author = {Ron, Dorit and Shamir, Adi},
  title = {{How Did Dread Pirate Roberts Acquire and Protect his Bitcoin Wealth?}},
  booktitle={Financial Cryptography and Data Security},
  year = {2014},
}

@inproceedings{huang2014botcoin,
  title = {{Botcoin: Monetizing Stolen Cycles}},
  author = {Huang, Danny Yuxing and Dharmdasani, Hitesh and Meiklejohn, Sarah and Dave, Vacha and Grier, Chris and McCoy, Damon and Savage, Stefan and Weaver, Nicholas and Snoeren, Alex C and Levchenko, Kirill},
  booktitle = {Network and Distributed Systems Security Symposium},
  year = {2014}
}

@inproceedings{behindLiao,
  author = {Kevin Liao and Ziming Zhao and Adam Doup{\'e} and Gail-Joon Ahn},
  title = {{Behind Closed Doors: Measurement and Analysis of CryptoLocker
           Ransoms in Bitcoin}},
  booktitle = {APWG Symposium on Electronic Crime Research},
  month = {June},
  year = {2016},
}

@article{economicConti,
  author = {Mauro Conti and Ankit Gangwal and Sushmita Ruj},
  title = {{On the Economic Significance of Ransomware Campaigns: {A} Bitcoin
           Transactions Perspective}},
  journal = {Computers \& Security},
  volume = {79},
  pages = {162-189},
  year = {2018},
  issn = {0167-4048},
}

@inproceedings{trackingHuang,
  title = {{Tracking Ransomware End-to-end}},
  author = {Danny Yuxing Huang and Maxwell Matthaios Aliapoulios and Vector Guo Li and Luca Invernizzi and Kylie McRoberts and Elie Bursztein and Jonathan Levin and Kirill Levchenko and Alex C. Snoeren and Damon McCoy},
  booktitle = {IEEE Symposium on Security and Privacy},
  month = {May},
  year = {2018},
}

@article{ransomwareClouston,
  title = {{Ransomware Payments in the Bitcoin Ecosystem}},
  author = {Paquet-Clouston, Masarah and Haslhofer, Bernhard and Dupont, Benoit},
  journal = {Journal of Cybersecurity},
  volume = {5},
  number = {1},
  pages = {tyz003},
  year = {2019},
  publisher = {Oxford University Press},
}

@inproceedings{backpagePortnoff,
  author = {Portnoff, Rebecca S. and Huang, Danny Yuxing and Doerfler, Periwinkle and Afroz, Sadia and McCoy, Damon},
  title = {{Backpage and Bitcoin: Uncovering Human Traffickers}},
  booktitle = {ACM SIGKDD International Conference on Knowledge Discovery and Data Mining},
  year = {2017},
}

@inproceedings{ponziBartoletti,
  author = {Massimo Bartoletti and Barbara Pes and Sergio Serusi},
  title = {{Data Mining for Detecting Bitcoin Ponzi Schemes}},
  booktitle = {Crypto Valley Conference on Blockchain Technology},
  month = {June},
  year = {2018},
}

@inproceedings{spamsPaquetClouston,
  author = {Paquet-Clouston, Masarah and Romiti, Matteo and Haslhofer, Bernhard and Charvat, Thomas},
  title = {{Spams Meet Cryptocurrencies: Sextortion in the Bitcoin Ecosystem}},
  booktitle = {ACM Conference on Advances in Financial Technologies},
  year = {2019},
}

@inproceedings{cybercriminalLee,
  title = {{Cybercriminal Minds: An Investigative Study of Cryptocurrency Abuses in the Dark Web}},
  author = {Seunghyeon Lee and Changhoon Yoon and Heedo Kang and Yeonkeun Kim and Yongdae Kim and Dongsu Han and Sooel Son and Seungwon Shin},
  booktitle = {Network and Distributed Systems Security Symposium},
  year = {2019}
}

@inproceedings {kappos22peel,
  author = {George Kappos and Haaroon Yousaf and Rainer St{\"u}tz and Sofia Rollet and Bernhard Haslhofer and Sarah Meiklejohn},
  title = {{How to Peel a Million: Validating and Expanding Bitcoin Clusters}},
  booktitle = {USENIX Security Symposium},
  year = {2022},
}

@article{xia2020characterizing,
  title = {{Characterizing Cryptocurrency Exchange Scams}},
  author = {Xia, Pengcheng and Wang, Haoyu and Zhang, Bowen and Ji, Ru and Gao, Bingyu and Wu, Lei and Luo, Xiapu and Xu, Guoai},
  journal = {Computers \& Security},
  volume = {98},
  pages = {101993},
  year = {2020},
  publisher = {Elsevier}
}

@article{bartoletti2021cryptocurrency,
  title = {{Cryptocurrency Scams: Analysis and Perspectives}},
  author = {Massimo Bartoletti and Stefano Lande and Andrea Loddo and Livio Pompianu and Sergio Serusi},
  journal = {IEEE Access},
  volume = {9},
  pages = {148353--148373},
  year = {2021},
  publisher = {IEEE}
}

@inproceedings{tekiner2021sok,
  title = {{SoK: Cryptojacking Malware}},
  author = {Tekiner, Ege and Acar, Abbas and Uluagac, A Selcuk and Kirda, Engin and Selcuk, Ali Aydin},
  booktitle = {IEEE European Symposium on Security and Privacy},
  year = {2021},
}

@inproceedings{li2023giveaway,
  title = {{Double and Nothing: Understanding and Detecting Cryptocurrency Giveaway Scams}},
  author = {Li, Xigao and Yepuri, Anurag and Nikiforakis, Nick},
  booktitle = {Network and Distributed Systems Security Symposium},
  year = {2023},
}

@misc{iocsearcher,
  title = {iocsearcher},
  note = {\url{https://github.com/malicialab/iocsearcher}},
  year = {2025},
}

@inproceedings{vasek2015there,
  title = {{There’s No Free Lunch, Even Using Bitcoin: Tracking the Popularity and Profits of Virtual Currency Scams}},
  author = {Vasek, Marie and Moore, Tyler},
  booktitle = {Financial Cryptography and Data Security},
  year = {2015},
}

@article{oggier2020ego,
  title = {{An Ego Network Analysis of Sextortionists}},
  author = {Oggier, Fr{\'e}d{\'e}rique and Datta, Anwitaman and Phetsouvanh, Silivanxay},
  journal = {Social Network Analysis and Mining},
  volume = {10},
  pages = {1--14},
  year = {2020},
  publisher = {Springer}
}

@misc{huggingface,
  title = {{Hugging Face}},
  note = {\url{https://huggingface.co}},
  year = {2023}
}

@misc{huggingface-models,
  key = {huggingface-models},
  title = {{SentenceTransformers Pre-trained models}},
  note = {\url{https://www.sbert.net/docs/pretrained_models.html}},
  year = {2025}
}

@misc{all-mpnet-base-v2,
  title = {{Sentence Transformers: all-mpnet-base-v2}},
  note = {\url{https://huggingface.co/sentence-transformers/all-mpnet-base-v2}},
  year = {2025}
}

@misc{openai-pricing,
  title = {{OpenAI Pricing}},
  note = {\url{https://openai.com/api/pricing}},
  year = {2025}
}

@misc{langdetect,
  title = {{langdetect}},
  note = {\url{https://pypi.org/project/langdetect}},
  year = {2025}
}

@misc{wordcloud,
  title = {{word\_cloud}},
  note = {\url{https://github.com/amueller/word_cloud}},
  year = {2025}
}

@article{sanh2019distilbert,
  title ={DistilBERT, a distilled version of BERT: smaller, faster, cheaper and lighter},
  author = {Sanh, Victor and Debut, Lysandre and Chaumond, Julien and Wolf, Thomas},
  journal = {arXiv preprint arXiv:1910.01108},
  year = {2019}
}

@inproceedings{hdbscan,
  title = {{Density-Based Clustering Based on Hierarchical Density Estimates}},
  author = {Campello, Ricardo J. G. B. and Moulavi, Davoud and Sander, Joerg},
  booktitle = {Advances in Knowledge Discovery and Data Mining},
  publisher = {Springer Berlin Heidelberg},
  year = {2013},
}

@article{choi2022large,
  title = {{A Large-Scale Bitcoin Abuse Measurement and Clustering Analysis Utilizing Public Reports}},
  author = {Choi, Jinho and Kim, Jaehan and Song, Minkyoo and Kim, Hanna and Park, Nahyeon and Seo, Minjae and Jin, Youngjin and Shin, Seungwon},
  journal = {IEICE TRANSACTIONS on Information and Systems},
  volume = {105},
  number = {7},
  pages = {1296--1307},
  year = {2022},
  publisher = {The Institute of Electronics, Information and Communication Engineers}
}

@article{buil2022offending,
  title={Offending concentration on the internet: An exploratory analysis of bitcoin-related cybercrime},
  author={Buil-Gil, David and Salda{\~n}a-Taboada, Patricia},
  journal={Deviant Behavior},
  volume={43},
  number={12},
  pages={1453--1470},
  year={2022},
  publisher={Taylor \& Francis}
}

@inproceedings{rosenquist2024dark,
  title = {{On the Dark Side of the Coin: Characterizing Bitcoin Use for Illicit Activities}},
  author = {Rosenquist, Hampus and Hasselquist, David and Arlitt, Martin and Carlsson, Niklas},
  booktitle = {International Conference on Passive and Active Network Measurement},
  year={2024},
}

@article{cross2023romance,
  title = {Romance baiting, cryptorom and ‘pig butchering’: an evolutionary step in romance fraud},
  author = {Cross, Cassandra},
  journal = {Current Issues in Criminal Justice},
  pages = {1--13},
  year = {2023},
  publisher = {Taylor \& Francis}
}

@article{ji2023survey,
  title = {{Survey of Hallucination in Natural Language Generation}},
  author = {Ji, Ziwei and Lee, Nayeon and Frieske, Rita and Yu, Tiezheng and Su, Dan and Xu, Yan and Ishii, Etsuko and Bang, Ye Jin and Madotto, Andrea and Fung, Pascale},
  journal = {ACM Computing Surveys},
  volume = {55},
  number = {12},
  pages = {1--38},
  year = {2023},
  publisher = {ACM New York, NY}
}

@misc{bbb_annual_2023,
  title={{BBB Scam Tracker} Risk Report},
  author={{Better Business Bureau Institute for Marketplace Trust}},
  month={April},
  year={2024},
  note = {\url{https://bbbmarketplacetrust.org/wp-content/uploads/2024/04/2023-BBBScamTracker-RiskReport-US-040224.pdf}},
}

@article{ftc2023report,
  title={Consumer Sentinel Network - Data Book 2023},
  author={{Federal Trade Commission}},
  month={February},
  year={2024},
  note = {\url{https://www.ftc.gov/system/files/ftc_gov/pdf/CSN-Annual-Data-Book-2023.pdf}},
  urldate = {2024-04-24},
}

@misc{llama3-huggingface,
  title={Meta-Llama-3-70B-Instruct},
  year={2024},
  note = {\url{https://huggingface.co/meta-llama/Meta-Llama-3-70B-Instruct}},
}

@article{tiwari2020future,
  title = {{The future of raising finance-a new opportunity to commit fraud: A review of Initial Coin Offering (ICOs) scams}},
  author = {Tiwari, Milind and Gepp, Adrian and Kumar, Kuldeep},
  journal = {Crime, Law and Social Change},
  volume = {73},
  pages = {417--441},
  year = {2020},
  publisher = {Springer}
}

@article{smith2007consumer,
  title = {Consumer scams in {A}ustralia: An overview.},
  author = {Smith, Russell G},
  journal = {Trends \& Issues in Crime \& Criminal Justice},
  number = {331},
  year = {2007}
}

@misc{glasschain,
  key = {glasschain},
  title = {{GlassChain}},
  note = {\url{https://glasschain.org}},
  year = {2025},
}

@misc{crystalblockchain,
  title = {{Crystal Lite}},
  note = {\url{https://lite.crystalintelligence.com}},
  year = {2025},
}

@misc{arkham,
  key = {arkham},
  title = {{Arkham Intel}},
  note = {\url{https://intel.arkm.com}},
  year = {2025},
}

@misc{baAcquisition,
  title = {{TRM Labs Acquires Bitcoinabuse.com}},
  note = {\url{https://www.globenewswire.com/en/news-release/2023/10/11/2758400/0/en/TRM-Labs-Acquires-Bitcoinabuse-com.html}},
  month = {October},
  year = {2023},
}

@inproceedings{bitaab2023beyond,
  title = {{Beyond Phish: Toward Detecting Fraudulent e-Commerce Websites at Scale}},
  author = {Bitaab, Marzieh and Cho, Haehyun and Oest, Adam and Lyu, Zhuoer and Wang, Wei and Abraham, Jorij and Wang, Ruoyu and Bao, Tiffany and Shoshitaishvili, Yan and Doup{\'e}, Adam},
  booktitle = {IEEE Symposium on Security and Privacy},
  year={2023},
}

@inproceedings{kotzias2023scamdog,
  title = {{Scamdog Millionaire: Detecting E-commerce Scams in the Wild}},
  author = {Kotzias, Platon and Roundy, Kevin and Pachilakis, Michalis and Sanchez-Rola, Iskander and Bilge, Leyla},
  booktitle = {Annual Computer Security Applications Conference},
  year = {2023}
}

@inproceedings{kotzias2025ctrl,
  author = {Platon Kotzias and Michalis Pachilakis and Javier {Aldana Iuit} and Juan Caballero and Iskander Sanchez-Rola and Leyla Bilge},
  title = {{Ctrl+Alt+Deceive: Quantifying User Exposure to Online Scams}},
  booktitle = {Network and Distributed Systems Security Symposium},
  year = {2025},
}

@misc{tmj4_scam_reporting,
  title ={Victim blaming prevents consumers from coming forward and reporting fraud},
  author = {TMJ4},
  note = {\url{https://www.tmj4.com/news/i-team/victim-blaming-prevents-consumers-from-coming-forward-and-reporting-fraud}},
  year = {2022},
}

@misc{citizens_advice_toolkit,
  title = {How to communicate about scams in an effective and engaging way},
  author = {{Citizens Advice}},
  note = {\url{https://www.cas.org.uk/system/files/citizens_advice_scams_awareness_toolkit2018b.pdf}},
  year = {2018},
}

@inproceedings{acharya2024conning,
  title = {{Conning the Crypto Conman: End-to-End Analysis of Cryptocurrency-based Technical Support Scams}},
  author = {Acharya, Bhupendra and Saad, Muhammad and Cin{\`a}, Antonio Emanuele and Sch{\"o}nherr, Lea and Nguyen, Hoang Dai and Oest, Adam and Vadrevu, Phani and Holz, Thorsten},
  booktitle = {IEEE Symposium on Security and Privacy},
  year = {2024}
}

@article{badawi2020cryptocurrencies,
  title = {{Cryptocurrencies Emerging Threats and Defensive
Mechanisms: A Systematic Literature Review}},
  author = {Badawi, Emad and Jourdan, Guy-Vincent},
  journal = {IEEE Access},
  volume = {8},
  pages = {200021--200037},
  year = {2020},
  publisher = {IEEE}
}

@misc{bsa,
  title = {{Bank Secrecy Act (BSA)}},
  author = {{Office of the Comptroller of the Currency}},
  note = {\url{https://www.occ.treas.gov/topics/supervision-and-examination/bsa/index-bsa.html}},
  year = {2025},
}

@misc{5amld,
  title = {{Directive (EU) 2018/843}},
  author = {{European Parliament}},
  note = {\url{https://eur-lex.europa.eu/eli/dir/2018/843/oj/eng}},
  month = {May},
  year = {2018},
}

@inproceedings{serverransom,
  author = {Kevin van Liebergen and Gibran Gomez and Srdjan Matic and Juan Caballero},
  title = {{All your (data)base are belong to us: Characterizing Database Ransom(ware) Attacks}},
  booktitle = {Network and Distributed Systems Security Symposium},
  year = {2025},
}

@misc{fbi,
  title = {{As Feds Fumble With Bitcoin, The Internet Trolls The FBI’s "Private" Wallet}},
  author = {{John Biggs}},
  note = {\url{https://techcrunch.com/2013/10/07/as-feds-fumble-with-bitcoin-the-internet-trolls-the-fbis-private-wallet/}},
  month = {October},
  year = {2013},
}

@misc{bitfinex,
  title = {{Bitfinex Github account}},
  author = {{bitfinexcom}},
  note = {\url{https://github.com/bitfinexcom/pub/blob/main/wallets.txt}},
  month = {December},
  year = {2022},
}

@misc{bitmex,
  title = {{Bitmex Proof of reserves}},
  author = {{Bitmex}},
  note = {\url{https://s3-eu-west-1.amazonaws.com/public.bitmex.com/data/porl/20221115-reserves-763269-20221115D113036434534000.yaml}},
  month = {November},
  year = {2022},
}

@misc{blockchair,
  key = {Blockchair},
  note = {\url{https://blockchair.com}},
  title = {{Blockchair}},
  year = {2025},
}

@misc{blockchain-com,
  key = {Blockchain-com},
  note = {\url{https://blockchain.com}},
  title = {{Blockchain.com}},
  year = {2025},
}

@misc{etherscan,
  key = {Etherscan},
  note = {\url{https://etherscan.io}},
  title = {{Etherscan}},
  year = {2025},
}

@misc{gpt-oss,
  key = {gpt-oss},
  note = {\url{https://openai.com/index/introducing-gpt-oss/}},
  title = {{Introducing gpt-oss}},
  month = {August},
  year = {2025},
}

\newpage
\appendix
\section{Additional Results}

This appendix provides additional tables and figures.
Figure~\ref{fig:query2} shows the prompt used in our naive LLM usage baseline,
which simply provides the LLM with the list of abuse types and requests 
the model to choose one.
Table~\ref{tab:bbb-estimation} details the comparison of the two revenue 
estimation approaches: using victim-reported losses and 
using address deposits.
The estimations are compared on 96 Bitcoin addresses appearing in 
71 Scam Tracker reports, as detailed in Section~\ref{sec:tagging}.
Finally, Table~\ref{tab:definitions} details the definitions used
by the LLM-based classifier.

\begin{figure}
\small
\centering
\begin{query}[Basic Prompt]
You are a cybersecurity expert with extensive knowledge about scams and abuses. You will help me classify abuse reports given by users, based exclusively on the content of the given TEXT. Do not infer or assume facts that are not described in the TEXT.

TEXT: $<$report\_description\_text$>$

QUESTION: Classify the given TEXT in one of the following abuse classes: $<$class\_list$>$. Answer only with the name of the class that clearly matches one of these classses and justify your answer by filling the next JSON structure: \{"answer": "", "reasoning": "" \}

\end{query}
\caption{LLM prompt used in the baseline. Macros in angle brackets 
are replaced with a list of abuse types and the description text.
Abuse type definitions are not included.}
\label{fig:query2}
\end{figure}

\begin{table}
\centering
\resizebox{\columnwidth}{!}{
\begin{tabular}{lrr|r|rr}
\hline
	\textbf{Abuse} & \textbf{Rep.} & \textbf{Addr.} & \textbf{Rep. Loss} & \textbf{BTC} & \textbf{USD} \\
\hline

        unauthwithdrawal & 1 & 1 & \$76,000 & 1,321.1525 & \$6,608,765 \\
        investment & 27 & 38 & \$129,160 & 192.8219 & \$6,479,591 \\
        advancefee & 5 & 20 & \$233,455 & 30.7403 & \$1,102,368 \\
        scam & 6 & 6 & \$8,360 & 19.0552 & \$767,115 \\
        giveaway & 5 & 5 & \$73,090 & 4.4790 & \$177,599 \\
        notabuse & 3 & 3 & \$300 & 1.8235 & \$28,414 \\
        reputation & 1 & 1 & - & 0.3088 & \$5,907 \\
        sextortion & 20 & 19 & - & 1.1376 & \$5,634 \\
        extortion & 2 & 2 & - & - & - \\
        abuse & 1 & 1 & - & - & - \\
\hline
	All & 71 & 96 & \$520,365 & 1,571.5188 & \$15,175,393 \\

\hline
\end{tabular}
}
\caption{For 71 Scam Tracker reports mentioning 96 Bitcoin addresses, 
comparison of victim-reported losses (Rep. Loss) with revenue 
computed from the address deposits. }
\label{tab:bbb-estimation}
\end{table}

\clearpage
\onecolumn

\begin{longtable}{|p{.15\linewidth}|p{.65\linewidth}|}
  \caption{List of definitions for the LLM}
  \label{tab:definitions} \\
\hline
\textbf{Category} & \textbf{Definition} \\
  \endhead
\hline
	abuse & An abuse report details instances of abuse or attempted abuse experienced by the person who authors the report, such as extortion, scams, demands for ransom, unauthorized withdrawals, replacement of addresses in the clipboard when copying/pasting cryptocurrency addresses, or theft. The report must point to how the abuse happend by providing a description or presenting evidence of the occurrence. It never promotes or recomend services of any kind. \\
	notabuse & A report is notabuse if it does not describe an abuse attempt suffered by the reporter but rather presents a personal opinion, calls for action, advertises a service, claims that an address is involved in terrorism, or that it belongs to hackers. A notabuse report may initially describe an abuse to look more convincent, but its main message always turns to be promoting something like a service, a personal opinion, a call for action, a claim, or a defamation. In general, a notabuse report may sound phishy or seem like a scam. \\
	scam & A scam is an attempt to defraud a person or group by gaining their confidence. Scammers are dishonest and deceive their victims to gain some benefit at the expense of the victim. Scammers may lie to the victims with texts that look legitimate but are hard to verify, commonly exploiting trust, compassion, greed, vulnerability, or lack of awareness in their victims. Popular scams involve investment scams, romance scams, giveaway scams, fake services like e-shops or funds recovery, and requests for donations to unverified addresses. \\
	extortion & Extortion is the practice of obtaining benefit through coercion. The victim is threatened with something bad happening in the future if it does not make a payment to the extortionist. The threat may involve revealing compromising data from the victim, the victim suffering physical harm, or the extortionist damaging assets belonging to the victims. To make the threat more credible the extortionist may claim to know some password of the victim. \\
	unauthwithdrawal & Unauthorized withdrawal refers to a third party withdrawing or transferring funds belonging to the victim without proper authorization or consent by the victim. Unauthorized withdrawal includes theft through credential stealing and the use of clipper malware to replace addresses copied into the clipboard with addresses belonging to the attacker when they are pasted. \\
	ransom & Attacks that demand a ransom payment from the victim in order to recover access to data encrypted or deleted by the attacker. The data may have been stored in a device or server owned by the victim. The attackers leave a ransom note explaining how to make the payment or how to contact the attackers for instructions. \\
	deleteransom & A ransom scheme where the attacker demands a ransom payment from the victim in order to recover data removed (and possibly exfiltrated) by the attacker, typically stored in databases. \\
	clipper & Unauthorized withdrawals enabled by a clipper. A clipper is a type of malware that replaces cryptocurrency addresses copied to the victim's clipboard with an address belonging to the attacker, without the victim being aware of the replacement. When the victim pastes the address copied to the clipboard as destination of a cryptocurrency transaction, the funds are unwillingly transferred to the attacker's address. \\
	bomb & A type of extortion where attackers threaten with detonating an explosive device. The attacker claims the bomb has been hidden in the building where the victim lives or works and that the explosion will affect many people. \\
	harm & A type of extortion where attackers threaten with physically harming the victim unless a payment is made. Beyond moral damage, the attacker threatens to physically hurt the victim or destroy its property. The victim is supposed to pay for the attacker to remain inactive. \\
	ddos & A type of extortion where attackers threaten with launching a distributed denial of service (DDoS) attack against a target server, website, or service. The DDoS attack will cause the service to be unavailable. The victim is requested to make a payment so that the DDoS attack does not happen. \\
	reputation & A type of extortion where attackers threaten with damaging the reputation of the victim by sending threats with insults to several recipients (for instance, other web site owners or bloggers) on the victim's behalf. \\
	investment & A type of scam that entices individuals to invest with the promise of high returns or guaranteed profits. Investment scams may use fake testimonials, fabricated success stories, or false celebrity endorsements to lure victims to invest. Scammers may create fake investment websites, platforms, or social media profiles to appear legitimate and trustworthy. After the initial investment, victims are not allowed to withdraw their profits or recover their original investment. Instead, attackers may request new fees and investments to allow withdrawal. \\
	advancefee & A type of scam where scammers promise the victim with a significant share of a large sum of money, in return for a small advance fee. The fraudster claims the up-front payment will be used to obtain the large sum. The scammer may impersonate a wealthy individual such as a Nigerian prince or an organization such as a tax revenue service or the social security administration. \\
	romance & A type of scam where victims are deceived by scammers to establish fake romantic relationships. Scammers create false personas, often online, and manipulate their victims' emotions to gain their trust and affection. Once an emotional bond is formed, the scammer demands payments from the victim under false excuses. \\
	eshop & A type of scam where scammers create an online shop website or e-commerce store where victims may place orders for buying items but receive nothing, or may receive a product that does not match the purchased one, e.g., a counterfeit or inferior product. The fake e-shop is made to look legitimate and lures buyers through unusually low prices or by offering illegal products. The e-shop may be offered as dark markets only accesible through Tor. \\
	fundsrecovery & A type of abuse report in which the author specifically mentions to being a victim of a fake funds-recovery service. The victim may describe how the financial losses occured, or how the perpetrator was contacted, but the main purpose of the report must always be to give details about the fake funds-recovery service or person, and never promote a service or person that recovers funds. \\
	sextortion & A type of extortion where attackers specifically claim to have a video or photos of the victim in sexual acts of any kind. The attackers may claim to have access to the victim's webcam. The attackers must threaten to share the video or photos with explicit sexual content with friends, family, or colleagues of the victim unless a payment is made. The report is not sextortion if it does not clearly mention sex-related material. \\
	giveaway & A type of scam where scammers promise that upon reception of a cryptocurrency deposit from the victim they will return to the sending address twice or more the received cryptocurrency amount. The scammers often lure victims to participate in the giveaway scam by impersonating famous persons, organizations, or cryptocurrency influencers, or by advertising the giveaway as a promotion or contest. \\
	employment & Employment scams advertise fake job opportunities or services for assisting individuals in finding employment. They are also known as job scams. Victims of employment scams typically do not receive a payment for the tasks they completed, receive only partial payment for their work, or are required to pay a fee to be able to withdraw their earnings. The offered jobs are often shady such as writing fake reviews, generating traffic to websites, money-laundering, or reshipping packages. \\
\hline
\end{longtable}

\clearpage
\twocolumn

\end{document}